\documentclass[lettersize,journal]{IEEEtran}
\pdfoutput=1
\usepackage{amsmath,amsfonts}
\usepackage{algorithmic}
\usepackage{algorithm}
\usepackage{array}
\usepackage[caption=false,font=normalsize,labelfont=sf,textfont=sf]{subfig}
\usepackage{textcomp}
\usepackage{stfloats}
\usepackage{url}
\usepackage{verbatim}
\usepackage{graphicx}
\usepackage{cite}
\usepackage{multirow}
\begin{document}

\title{BESA: Boosting Encoder Stealing Attack with Perturbation Recovery}

\author{Xuhao Ren, Haotian Liang, Yajie Wang, Chuan Zhang, \IEEEmembership{Member,~IEEE,} Zehui Xiong, \IEEEmembership{Senior Member, ~IEEE,} Liehuang Zhu, \IEEEmembership{Senior Member, ~IEEE}
\thanks{This work was financially supported by the National Natural Science Foundation of China (Grant Nos. 62472032, 62202051, and 62232002); the Open Project Funding of Key Laboratory of Mobile Application Innovation and Governance Technology, Ministry
of Industry and Information Technology (Grant No. 2023IFS080601-K); the Beijing Institute of Technology Research Fund Program for Young Scholars; the Young Elite Scientists Sponsorship Program by CAST (Grant No. 2023QNRC001); Xiaomi Research Fund for Young Scholars, and the Fundamental Research Funds for the Central Universities (Grant No. 2024CX06034); and the BIT Research and Innovation Promoting Project (Grant No. 2024YCXZ022).}
\thanks{Xuhao Ren, Haotian Liang, Yajie Wang, Chuan Zhang, and Liehuang Zhu are with the
School of Cyberspace Science and Technology, Beijing Institute of Technology, Beijing 100081, China (e-mail: xuhaor@bit.edu.cn; haotianl@bit.edu.cn; wangyajie0312@foxmail.com; chuanz@bit.edu.cn; liehuangz@bit.edu.cn).}
\IEEEcompsocitemizethanks{\IEEEcompsocthanksitem Zehui~Xiong is with Singapore University of Technology and Design, Singapore (email:~zehui\_xiong@sutd.edu.sg).}
\thanks{Corresponding author: Chuan Zhang.
}}

\markboth{Journal of \LaTeX\ Class Files,~Vol.~14, No.~8, August~2021}%
{Shell \MakeLowercase{\textit{et al.}}: A Sample Article Using IEEEtran.cls for IEEE Journals}


\maketitle

\begin{abstract}
To boost the encoder stealing attack under the perturbation-based defense that hinders the attack performance, we propose a boosting encoder stealing attack with perturbation recovery named BESA. It aims to overcome perturbation-based defenses. The core of BESA consists of two modules: perturbation detection and perturbation recovery, which can be combined with canonical encoder stealing attacks. The perturbation detection module utilizes the feature vectors obtained from the target encoder to infer the defense mechanism employed by the service provider. Once the defense mechanism is detected, the perturbation recovery module leverages the well-designed generative model to restore a clean feature vector from the perturbed one. Through extensive evaluations based on various datasets, we demonstrate that BESA significantly enhances the surrogate encoder accuracy of existing encoder stealing attacks by up to 24.63\% when facing state-of-the-art defenses and combinations of multiple defenses.
\end{abstract}

\begin{IEEEkeywords}
Encoder stealing attack, perturbation recovery, perturbation detection, generative model.
\end{IEEEkeywords}

\section{Introduction}
\label{sec:intro}

Pre-trained encoders are extensively utilized across various domains in real-world scenarios \cite{baevski2023efficient}. However, training well-performing pre-trained encoders is a time-consuming, resource-intensive, and costly process \cite{qu2023reaas}. Hence, encoder owners are highly motivated to safeguard the privacy of their pre-trained encoders.

Unfortunately, recent works have shown that pre-trained encoders are susceptible to encoder stealing attacks \cite{sha2023can}. These attacks allow an attacker to create a surrogate encoder that closely mimics the functionality of a targeted encoder by simply querying it through the APIs. 
The consequences of such attacks can be quite severe. On the one hand, many service providers, such as OpenAI, Google, and Meta, offer cloud-based Encoder as a Service (EaaS) solutions to monetize their pre-trained encoders \cite{zhang2023awencoder}. 
Users compensate these service providers for accessing the encoder through Application Programming Interfaces (APIs). 
However, an attacker can acquire the cloud-based encoder at a significantly reduced expense compared to the investment in data collection and training, which not only infringes on intellectual property but also results in financial setbacks for the original service provider \cite{jia202110,dziedzic2022difficulty,dong2023rai2}. 
On the other hand, encoder stealing attack can also act as a launchpad for various types of attacks like adversarial sample attacks \cite{chen2022self}, membership inference attacks \cite{liu2021encodermi,he2022semi}, and backdoor injection attacks \cite{jia2022badencoder,saha2022backdoor,li2023embarrassingly}. For example, certain adversarial example attacks rely on having access to the gradient of the target encoder, which is inaccessible in black-box scenarios. In such cases, an adversary can create a surrogate encoder through model encoder stealing attacks and generate adversarial examples using the white-box surrogate encoder. Additionally, researchers have demonstrated that a surrogate encoder obtained through encoder stealing attacks can enable membership inference attacks as well as more damaging backdoor injection attacks.

\begin{figure*}[htbp]
\centering
\includegraphics[width=0.9\textwidth]{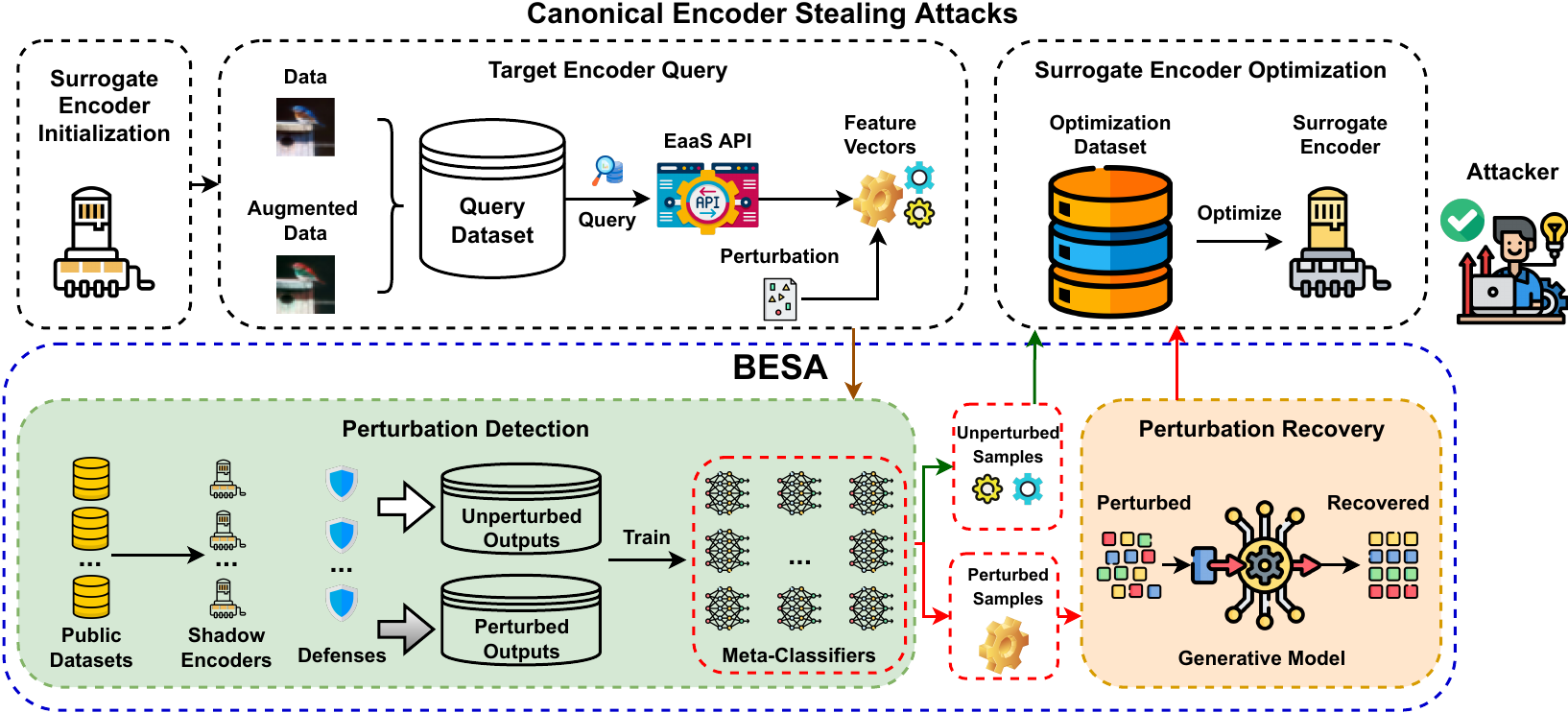} 
\caption{The architecture of BESA.}
\label{fig:Intro}
\end{figure*}

Extensive research has been conducted to mitigate encoder stealing attacks in various ways. Common defense strategies include detection methods \cite{juuti2019prada,tang2024modelguard}, watermarking techniques \cite{li2023plmmark,li2023ssl,tang2023watermarking}, and perturbation-based approaches \cite{marshall2019threat,liu2022stolenencoder}. 
Among these methods, the perturbation-based approaches have been widely adopted by many EaaS providers based on their performance in degrading existing encoder stealing attacks.
Based on their good performance in real-world scenarios, these defense strategies have been adopted by real-world EaaS providers to safeguard against encoder stealing attacks.
For example, Liu et al. \cite{liu2022stolenencoder} have demonstrated that using perturbation-based defense mechanisms can cause a decrease in the accuracy of the substitute encoder on downstream classification tasks, from 78.12\% to 42.07\%.
Therefore, it is necessary for the attackers to explore practical ways to bypass or penetrate such defense methods for more effective stealing attacks.

To the best of our knowledge, previous works have widely ignored the possibility of performing encoder stealing attacks against defended target encoders. 
Motivated by this research gap, we propose a boosting encoder stealing attack with perturbation recovery, called BESA. The core idea behind BESA is to detect and recover perturbed feature vectors, which is illustrated in Fig. \ref{fig:Intro}. To be specific, canonical encoder stealing attacks typically involve three steps. Initially, the surrogate encoder is either initialized as empty or pre-trained. Next, original or augmented samples \cite{liu2022stolenencoder} are chosen to query the target encoder for feature vectors. Subsequently, the surrogate encoder is optimized using these feature vectors. This query and optimization process is iterated until optimal performance is achieved. However, if the feature vectors are perturbed by the service provider, the optimization efforts may prove futile. As shown in Section \ref{sec:expr}, the performances of existing works are seriously degraded if the feature vectors are perturbed by the service provider.
To solve this problem, we introduce two general modules in BESA, namely perturbation detection and perturbation recovery, which can be incorporated into canonical encoder stealing attacks before the optimization phase for the surrogate encoder.

To achieve our objective of boosting encoder stealing attacks, we encounter several challenges as follows.

\begin{itemize}
    \item \textbf{Detecting the adopted defense method is challenging.}
\end{itemize}

Generally, the defense method adopted by the service provider is often unknown. 
Although some detecting methods have been proposed recently \cite{chen2023d}, they mainly aim at logits with small sizes and simple distribution characteristics. 
However, existing methods fall short as the feature vectors from pre-trained encoders are always large and with complex distribution characteristics.
To tackle this problem, we first train a collection of shadow encoders applied with multiple defenses. Here, the data augmentation technique is adopted to overcome the lack of enough public training data.
Subsequently, we train a binary meta-classifier with multiple layers for each defense method using perturbed and unperturbed feature vectors.
The trained meta-classifier can accurately predict if a particular feature vector has been altered by the respective defense technique. The results from the experiments in Section \ref{sec:expr-detection} demonstrate its ability to differentiate the defense method with an accuracy of over 99\%.

\begin{itemize}
    \item \textbf{Recovering perturbed feature vectors is challenging.}
\end{itemize}
    
Even if we can detect the defense method, the precise amount of noise added remains uncertain.
Although generative models have been used for perturbation recovery \cite{chen2023d}, their architectures are designed for logits with small sizes and simple distribution.
However, these models are inefficient and ineffective for feature vectors with large sizes and complex distribution.
To tackle this challenge, we construct a generative model inspired by the MagNet \cite{meng2017magnet} based on its outstanding performance of perturbation recovery on samples with large size and complex distribution. 
Next, we proceed to train the model using perturbed feature vectors as inputs and unperturbed ones as outputs. The experiments detailed in Section \ref{sec:expr-impr} illustrate that this method can enhance the accuracy of current attacks by as much as 24.63\%.

The contributions of this paper are as follows.

\begin{itemize}
    \item We have developed a boosting encoder stealing attack with perturbation recovery called BESA, illustrated in Fig. \ref{fig:Intro}, to enhance the effectiveness of canonical attacks against perturbation-based defenses.
    \item We have devised algorithms for perturbation detection and recovery, enabling the identification of the defense methods and the recovery of perturbed feature vectors for the optimization of the surrogate encoder.
    \item We have conducted experiments across different defense methods on three state-of-the-art attacks to evaluate the performance of BESA. The results show that it can improve the accuracy of existing attacks by up to 24.63\%.
\end{itemize}

The rest of the paper is structured as follows. Section \ref{sec:related} presents the related works. In Section \ref{sec:threat}, we give the threat model of our scheme, including the object, capabilities, and limitations. We give the scheme details in Section \ref{sec:detail}, followed by experiments in Section \ref{sec:expr}. Finally, we conclude this paper in Section \ref{sec:conclusion}.

\section{Related Works}
\label{sec:related}

\subsection{Model Stealing Attacks}
Most model stealing attacks mainly focus on the classifiers instead of the emerging pre-trained encoders \cite{gong2020model,gong2021inversenet}. 
For instance, Tramer et al. \cite{tramer2016stealing} were the first to investigate model stealing attacks in supervised learning. 
They demonstrated the feasibility of extracting the functionality of high-performing machine learning models deployed online through APIs.
Subsequently, extensive research has been conducted on various aspects \cite{yue2021black,ma2023divtheft,shah2023data,kariyappa2021maze,truong2021data,jia2021entangled,szyller2021good,zhou2020dast,lin2023quda}. 
For example, if the attacker manages to obtain similar or in-distribution data that resemble the surrogate dataset used for attacks, they can leverage data augmentation or active learning techniques by combining the datasets to query the API \cite{orekondy2019knockoff,papernot2017practical}. 
Furthermore, Truong et al. \cite{truong2021data} and Kariyappa et al. \cite{kariyappa2021maze} have also explored the possibility of stealing the model in data-free \cite{wang2022dst}.
However, this approach may prove insufficient in practical scenarios where the tasks involve significant commercial value, and the associated training dataset is considered highly confidential and hard to access \cite{shah2023data,ma2023divtheft}.

Numerous researchers have proposed data-free methods for extracting models to tackle this challenge \cite{zhou2020dast}. In such scenarios, the adversary lacks any information about the dataset used to train the target models.
Two recently developed techniques, namely DFME \cite{truong2021data} and MAZE \cite{kariyappa2021maze}, have been specifically designed to extract the functionality of target models under this challenge. However, these studies impose significant query budgets on the adversary, rendering them impractical in real-world scenarios. 
Lin et al. \cite{lin2023quda} tackled this problem by integrating Generative Adversarial Networks (GAN) to make use of weak image priors. They also employed deep reinforcement learning methods to enhance the query efficiency of data-free model extraction attacks.

In the aspect of emerging pre-trained encoders, Liu et al. \cite{liu2022stolenencoder} introduced a pioneering attack named ``StolenEncoder" with the objective of extracting the functionality of pre-trained encoders. Their method formulates the stealing attack as an optimization problem and utilizes the standard stochastic gradient descent paradigm to solve it. 
To optimize the query budget, the attack incorporates data augmentations to enhance its effectiveness. 
Additionally, a method called ``Cont-Steal" is introduced by Sha et al. \cite{sha2023can}, which employs the concept of contrastive learning to effectively utilize the rich information in the feature vectors. However, existing works on encoder stealing attacks have disregarded the exploration of stealing pre-trained encoders under perturbation-based defensive approaches. In contrast, our work focuses on encoder stealing attacks against defended pre-trained target encoders, which is more applicable in real-world scenarios.

\subsection{Defensive Approaches}

When protecting the pre-trained encoders against encoder stealing attacks, the service provider encounters two opposing objectives: impeding malicious and enhancing benign queries. 
In other words, the service provider aims to hinder the attacker's efforts in carrying out encoder stealing attempts while improving the performance of legitimate queries. 
Generally, existing defensive approaches in defending against encoder stealing attacks can be classified into three main categories: detection, watermarking, and perturbation. 

\subsubsection{Detection-based Methods.} Detection-based methods focus on determining if a query sequence is malicious, without modifying the feature vectors. 
In cases where a query sequence is flagged as malicious, the service provider might opt to adjust the feature vector or refuse service to the user. 
For the detection-based method, Dubi{\'n}ski et al. \cite{dubinski2023bucks} proposed an active defense method to counter encoder stealing attacks. 
This method prevents stealing in real-time, without compromising the quality of the representations for legitimate API users. 
Specifically, their defense approach is based on the observation that the representations provided to adversaries attempting to steal the encoder's functionality cover a significantly larger portion of the embedding space compared to representations of normal users using the encoder for a specific downstream task.
However, how to accurately define the concept of an ``anomalous query" still becomes challenging in practice.

\subsubsection{Watermarking-based Methods.} Watermarking-based methods involve injecting a carefully designed watermark backdoor into the target encoder.
This allows for effective transfer to the surrogate encoder of the attacker, aiding in copyright verification.
For the watermarking-based method, Cong et al. \cite{cong2022sslguard} introduced the first robust watermarking for pre-trained encoders called SSLGuard. 
This algorithm serves as a defense against model extraction and other watermark recovery attacks like input noising, output perturbing, overwriting, model pruning, and fine-tuning. 
Additionally, Peng et al. \cite{peng2023you} introduced EmbMarker, a backdoor-based watermarking scheme to protect pre-trained encoders in the large language aspect.
However, existing watermarking-based methods still face serious threats to extraction-based attacks due to their functional-irrelevant characteristic.
known for its ability to protect the copyright of pre-trained encoders through the use of Backdoor Watermark.

\subsubsection{Perturbation-based Methods}
The perturbation-based methods perturb the feature vectors of some or all queries. 
For the perturbation-based method, Liu et al. \cite{liu2022stolenencoder} first discuss perturbation-based methods for defending against their proposed \textit{StolenEncoder} attack on pre-trained encoders.
According to their categorization, there are three main approaches adopted in defending the pre-trained encoders. Firstly, the top-$k$ features approach resets the contents that are not among the top $k$ largest absolute values to 0. Secondly, the feature rounding approach rounded feature vectors. Lastly, the feature poisoning approach adds carefully crafted perturbations to the feature vector.
Nevertheless, the trade-off between the security under perturbation-based defenses and the usability for regular users still warrants further discussion.
According to their categorization, three main approaches have been adopted to defend the pre-trained encoders. Firstly, in the top-$k$ features approach, the EaaS API resets the contents of features that are not among the top $k$ largest absolute values to 0 before returning a feature vector to a customer. Secondly, in the feature rounding approach, the EaaS API returns rounded feature vectors to a customer. Lastly, the feature poisoning approach involves the EaaS API adding carefully crafted perturbations to a feature vector in order to manipulate the optimization of the surrogate encoder.

\section{Threat Model}
\label{sec:threat}
In this section, we establish the threat model based on the objective, capabilities, and the attack's 
limited knowledge.

\subsection{Objective}


In BESA, the attacker's goal is to create a surrogate encoder that imitates the actions of the target encoder, which is safeguarded by specific defense mechanisms. Additionally, the attacker aims to achieve this objective while working within restricted query budgets.

\subsection{Capabilities}
In BESA, we consider the attacker to possess three specific capabilities during the attack.

\begin{itemize}
    \item{\textbf{The attacker has access to publicly available datasets.} The attacker has the ability to gather datasets that are either consistent or inconsistent with the original training distribution of the target encoder. For instance, if the target encoder is pre-trained on the CIFAR-10 \cite{krizhevsky2009learning} dataset for downstream tasks, the CIFAR-100 dataset could be considered an in-distribution dataset, while the SVHN \cite{yuval2011reading} dataset, which consists of Google Street View house numbers, could be considered an out-of-distribution dataset.}
    
    \item{\textbf{The attacker has the capability to make queries to the target encoder.}
 The attacker interacts with the target encoder by making queries through the EaaS API or other interfaces to retrieve the relevant query results. It is assumed that the query results consist of feature vectors, which is a commonly found feature in most EaaS systems. However, due to the limited query budget, the attacker is only able to make a restricted number of queries to the target encoder. In the case of BESA, it is assumed that service providers utilize a defense strategy based on perturbation to modify some or all of the feature vector outcomes while preserving accuracy or adhering to accuracy constraints.}
    
    \item{\textbf{The attacker has the ability to reconstruct the defensive strategy.}
    The attacker has the ability to reconstruct a set of defense strategies labeled as ${f_{De} = \{f^{1}, f^{2}, \cdots, f^{K}\}}$, where each $f^{k},\ k \in \{1, 2, \cdots, K\}$ represents a specific defense tactic. It is assumed that the strategy employed by the service provider is included in this reconstructed set $f_{De}$. Importantly, we demonstrate the effectiveness of BESA through our experiments by showing its effectiveness even when the service provider adopts defense strategies that are not included in the attacker's reconstructed set.}
\end{itemize}

\subsection{Limitations}
In BESA, we make assumptions about the attacker's limitations. First, the attacker lacks knowledge about the target encoder, including its architecture, parameters, and hyperparameters. Moreover, it cannot access the original pre-trained samples of the target encoder due to their unavailability, difficulty in obtaining them, or prohibitively high cost. Finally, it lacks knowledge of the specific defense strategies employed by the service provider. The service provider is capable of selectively altering a subset of feature vectors, for which the attacker lacks detailed information. The details are as follows.
\begin{itemize}
    \item \textbf{No information about the Target Encoder.} The attacker lacks knowledge about the target encoder, including its architecture, parameters, and hyperparameters.
    \item \textbf{No Access to Original Pre-train Dataset.} The attacker cannot access the original pre-trained samples of the target encoder due to their unavailability, difficulty in obtaining them, or prohibitively high cost.
    \item \textbf{Inadequate Defense Strategy Knowledge.} The attacker lacks knowledge of the specific defense strategies employed by the service provider. The service provider is capable of selectively altering a subset of feature vectors, for which the attacker lacks detailed information.
\end{itemize}

\section{Detailed Construction}
\label{sec:detail}

In contrast to traditional encoder stealing attacks, BESA is improved by incorporating two additional modules positioned between the target encoder query module and the surrogate encoder optimizing module. These additional modules are intended to identify and restore perturbation feature vectors, as demonstrated in Fig. \ref{fig:Intro}. The first module, the perturbation detection module, utilizes meta-classifiers to recognize the defense mechanisms implemented through the service provider. Another module, the perturbation recovery module, focuses on recovering pristine feature vectors from the perturbed ones. These two modules can be seamlessly integrated into existing canonical encoder-stealing attack frameworks. This section provides a comprehensive overview of both the perturbation detection module and the perturbation recovery module and we present the detailed procedure of BESA in Algorithm \ref{alg:algorithm}.

\begin{algorithm}[tb]
\caption{BESA: Boosting encoder stealing attack with perturbation recovery}
\label{alg:algorithm}
\begin{algorithmic}[1] 
\REQUIRE API of the target encoder $\mathcal{T}$, public datasets with data augmentation $D_{1}, D_{2}, \cdots D_{M}$, the series of defense tattics $f^{1}, f^{2}, \cdots f^{K}$.
\ENSURE Trained surrogate encoder $\mathcal{S}$
\STATE // Preparation.
\STATE Train shadow encoders $E_{1}, E_{2}, \cdots E_{M}$ using the datasets $D_{1}, D_{2}, \cdots D_{M}$.
\FOR{$k \in [1, K]$}
\STATE Leverage $f^{k}$ to $E_{m},\ m \in [1, M]$.
\STATE Use the inputs $(x_{m}, E_{m}^{k}(x_{m})),\ x_{m} \in D_{m},\ \forall m$ as positive samples and $(x_{m}, E_{m}(x_{m})),\ \forall m$ as negative samples to train a binary meta-classifier $B_{k}$, where $E^k_m(x_m)$ is the output of encoder $E_m$ under defense $f^k$.
\STATE  Train a generative model $G_{k}$ using  $E_{m}^{k}(x_{m})$ as inputs and $E_{m}(x_{m})$ as outputs for all $m$.
\ENDFOR
\STATE // Surrogate encoder initialization.
\STATE Initialize a raw or a pre-trained surrogate encoder $\mathcal{S}$.
\WHILE{The query budget is not exhausted or $\mathcal{S}$ is not converged as desired}
\STATE // Target encoder query.
\STATE Construct a query sample $x_{q}$.
\STATE Use $x_{q}$ to query the target encoder $\mathcal{T}$ and receive the feature vector $\mathcal{T}(x_{q})$.
\STATE // Perturbation detection.
\STATE Input $(x_{q}$, $\mathcal{T}(x_{q}))$ into $B_{1}, B_{2}, \cdots B_{K}$.
\IF{There exist available prediction results}
\STATE Choose the prediction result with the highest confidence score, which is represented as $k^{*}$.
\STATE // Perturbation recovery.
\STATE Input $\mathcal{T}(x_{q})$ into $G_{k^{*}}$ to get $\mathcal{T}(x_{q}) \leftarrow G_{k^{*}}(\mathcal{T}(x_{q}))$.
\ENDIF
\STATE // Surrogate encoder optimizing.
\STATE Use $x_{q}$ and the recovered $\mathcal{T}(x_{q})$ to optimize the surrogate encoder $\mathcal{S}$.
\ENDWHILE

\end{algorithmic}
\end{algorithm}

\subsection{Perturbation Detection}

Our main objective is to identify the defense strategies employed by the service provider by leveraging the distinct characteristics of perturbed feature vectors associated with different defense strategies. To achieve this, we start by constructing a pool of unprotected pre-trained encoders. These encoders, denoted as $\{E_{1}, E_{2}, \cdots, E_{M}\}$, are trained on various publicly available datasets $\{D_{1}, D_{2}, \cdots, D_{M}\}$ using different architectures and contrastive learning algorithms. To construct more training data samples for shadow encoders with limited public datasets, the attacker can adopt the data augmentation technique utilized in various works \cite{liu2022stolenencoder,sha2023can}. Note that $E_m,\ m \in [1, M]$ refers to a set of unprotected encoders pre-trained on $D_m$ since we utilize different architectures and contrastive learning algorithms to train multiple models for each public dataset. Although it would be ideal for these shadow encoders to have architectures similar to those of the target encoder, this is not possible in black-box scenarios. \cite{liu2022stolenencoder}. Thus, we incorporate diversity into the designs of the shadow encoders to enhance overall generalization. Following this, we implement a range of defense strategies for these shadow encoders. With $K$ defense techniques available and $M$ groups of shadow encoders, we have the flexibility to assign a particular defense approach to multiple shadow encoders or opt for employing multiple defense methods for a single group of shadow encoders. Our goal is to secure an ample number of pairs comprising protected and unprotected shadow encoders for each defense tactic.

Here, we define $E_{m}^{k}$ as the protected version of $E_{m}$ under the defense strategy $f^{k}$. $E_m^k(x_m)$ denotes the output feature vector, while $E_m(x_m)$ is the corresponding clean output from the same encoder without any defense. These notations help distinguish between perturbed and unperturbed features, which are used to train the meta-classifiers. We expect that $E_{m}(x_{m}),\ x_{m} \in D_{m}$ captures the characteristics of unperturbed feature vectors, while $E_{m}^{k},\ x_{m} \in D_{m}$ embodies the characteristics of feature vectors using defense strategy $f^{k}$. Building on this observation, we train $K$ binary meta-classifiers, $B_{1}, \cdots, B_{K}$, where $B_{k}$ is responsible for detecting whether the $k$-th defense approach safeguards the feature vectors. Positive samples $(x_{m}, E_{m}^{k}(x_{m})),\ \forall m$ (with the label is $1$) and negative samples $(x_{m}, E_{m}(x_{m})),\ \forall m$ (with the label is $0$) are utilized in the training of the corresponding meta-classifier $B_{k}$. Since $B_{k}$ is being trained with samples from various classification tasks, it is anticipated that the model will grasp the inherent distinctions in characteristics between modified and unaltered feature vectors resulting from the defense mechanism $f^{k}$. This enables the meta-classifiers to generalize to the target encoder, even when the training dataset of the target encoder is unknown to the adversary.

Trained meta-classifiers are employed to determine the defense tactics used by the target encoder. Set $x_{q} \in D_{q}$ represent a query sample from the query dataset, and $\mathcal{T}(x_{q})$ represents the feature vector. The pair $(x_{q}, \mathcal{T}(x_{q}))$ is input into the meta-classifiers $B_{1}, \cdots B_{K}$ to obtain the corresponding prediction results. If all predictions are below a specified threshold (e.g., 0.5), the sample is unperturbed. However, if any prediction result exceeds the threshold $T_{h}$, we compare the confidence scores of these predictions and select the one with the highest confidence as the predicted defense tactic. Our experiments demonstrate that in the majority of cases, the meta-classifier associated with the actual defense strategy generates a confidence score exceeding 90\%.

\subsection{Perturbation Recovery}

If the feature vector of $x_q$ is predicted to be intact, we will optimize the surrogate encoder directly using this feature vector. However, if it is disrupted, we will retrieve the original feature vector from the altered one using the specified defense strategy. It is crucial to highlight that defense detection can be conducted at various intervals, including one-time, regular intervals, or for each query. In the most precise scenario, if the service provider alternates between different defenses randomly, the attacker can perform defense detection on every query outcome and subsequently restore each altered feature vector based on the detection outcomes.

To retrieve the feature vector safeguarded through a particular defense strategy $f^{k}$, we construct a generative model $G_{k}$ and compute the perturbed feature vector $\mathcal{T}(x_{q})$ as follows:
\begin{equation}
    \mathcal{T}(x_{q}) \gets G_k(\mathcal{T}(x_{q}))
\end{equation}
This model takes $\mathcal{T}(x_{q})$ as input and produces the corresponding unperturbed feature vector. In order to train the generative model $G_{k}$, we utilize the shadow encoders initially set up to detect perturbations in the previous step. To be specific, each training sample follows the format $(E_{m}^{k}(x_{m}), E_{m}(x_{m})),\ \forall m$. Here, $E_{m}^{k}(x_{m})$ represents the protected feature vector by defense strategy $f^{k}$ for shadow encoder $E_{m}$, while $E_{m}(x_{m})$ refers to the corresponding clean feature vector. By using this approach, $G_{k}$ acquires the ability to map perturbed feature vectors to unperturbed ones, thereby ensuring that $G_{k}(E_{m}^{k})$ exhibits the same characteristics as $E_{m}$.

\begin{figure}[htbp]
\centering
\includegraphics[width=1\columnwidth]{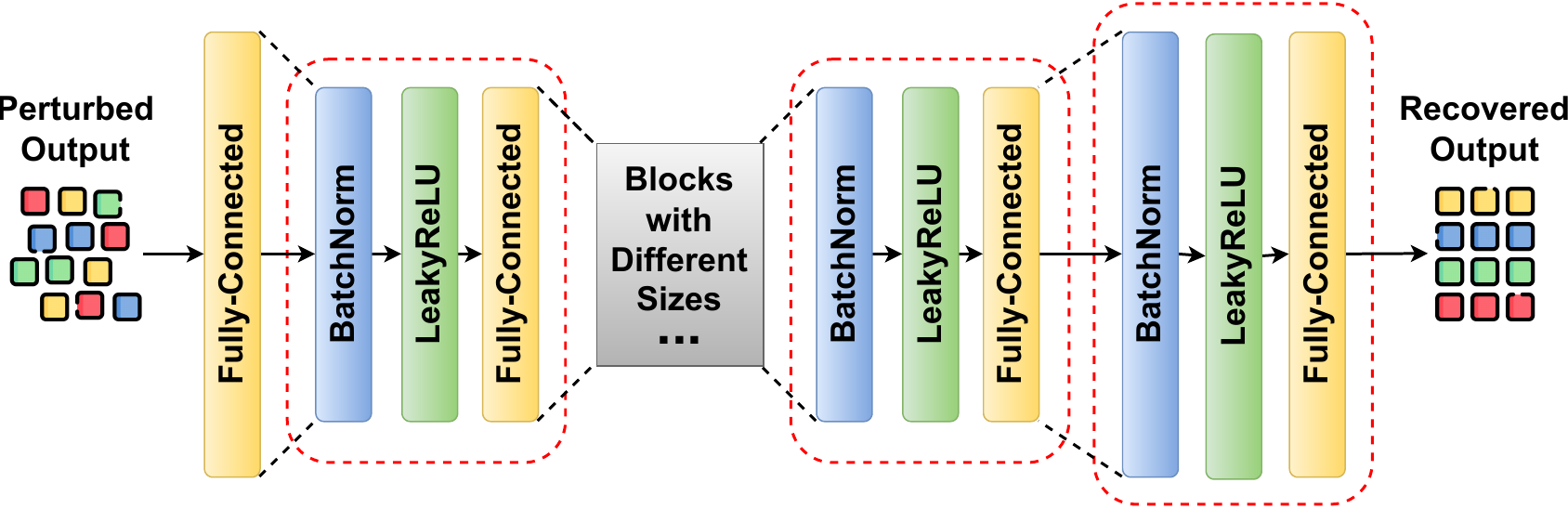} 
\caption{The design of the generator in BESA}
\label{fig:MagNet}
\end{figure}

Two major families of generative models are commonly used in the field: Variational Autoencoders \cite{kingma2013auto} and Generative Adversarial Networks (GANs) \cite{goodfellow2014generative}. Autoencoders are primarily employed for dimensionality reduction, compressing data samples into informative representations with smaller dimensions than the input. On the other hand, GANs are typically used for inverse transformations, generating samples of more complex distributions based on random noise with larger dimensions than the input. However, in the case of perturbation discovery, the dimensionality of the clean feature vector remains the same as that of the perturbed feature vector. Therefore, neither VAEs nor GANs are suitable for achieving the objective of perturbation recovery.

In Fig. \ref{fig:MagNet}, we have developed a generator that draws inspiration from MagNet \cite{meng2017magnet}. This generator denoted as $G$, is composed of multiple blocks comprising BatchNorm, LeakyReLU, and fully connected layers. Within each block, the BatchNorm layer initializes by normalizing the feature vectors to address convergence difficulties and mode collapse. Subsequently, the LeakyReLU activation function is employed rather than ReLU to alleviate the issue of a vanishing gradient \cite{maas2013rectifier}. Finally, the perturbed feature vectors are processed by the fully connected layer to eliminate the influence of defense strategies and reconstruct the corresponding clean feature vectors. Our proposed generator is designed to be lightweight but contains all essential components for effective perturbation recovery.

The generator is trained to minimize the difference between the recovered feature vector $G_{k}(E_{m}^{k}(x_{m}))$ and the actual clean feature vector $E_{m}(x_{m})$. The minimization optimization problem can be formulated as follows:
\begin{equation}
    \label{equ:aim}
   \underset{\theta_{G_{k}}}{\arg \min} \mathbb{E}_{x_{m} \sim \mathcal{D}_{m}}[\mathcal{L}(G_{k}(E_{m}^{k}(x_{m})), E_{m}(x_{m}))],
\end{equation}
where $\theta_{G_{k}}$ are the parameters of the generative model $G_{k}$, and $\mathcal{L}$ represents the loss function. Commonly employed loss functions in this context are cosine similarity, $L_{2}$-norm loss, and $L_{1}$-norm loss.

\textbf{Cosine Similarity.} Cosine similarity is a commonly used loss function in various encoder training algorithms and encoder stealing attacks. It provides an effective measure of the dissimilarity between two distinct feature vectors produced by their respective encoders.

\begin{equation}
    \label{equ:cos-loss}
   L_{C} = - \frac{1}{|\sum_{m} D_{m}|}\sum_{x_{m} \in D_{m}} \frac{G_{k}(E_{m}^{k}(x_{m})) \cdot E_{m}(x_{m})}{||G_{k}(E_{m}^{k}(x_{m}))|| \times ||E_{m}(x_{m})||}.
\end{equation}

\textbf{$L_{2}$-norm loss.} The $L_{2}$-norm loss is a straightforward loss function known for its rapid convergence.

\begin{equation}
    \label{equ:l2-loss}
   L_{2} = \frac{1}{|\sum_{m} D_{m}|}\sum_{x_{m} \in D_{m}}[G_{k}(E_{m}^{k}(x_{m})) - E_{m}(x_{m})]^{2}.
\end{equation}

\textbf{$L_{1}$-norm loss.} The $L_{1}$-norm loss exhibits a slower and less substantial gradient decay compared to the $L_{2}$-norm loss, but the derivative of the $L_{1}$-norm loss is not unique at the point 0, potentially leading to non-convergence.

\begin{equation}
    \label{equ:l1-loss}
   L_{1} = \frac{1}{|\sum_{m} D_{m}|}\sum_{x_{m} \in D_{m}}[G_{k}(E_{m}^{k}(x_{m})) - E_{m}(x_{m})].
\end{equation}

\begin{table*}[tb]
\centering
\caption{Accuracy of the surrogate encoder for different datasets. The higher level of accuracy is emphasized in bold.}
\label{tab:enhance}
\resizebox{13cm}{!}{%
\begin{tabular}{cc|cccccc}
\hline
Dataset                        & Method          & Top-K            & RD               & NP               & Hybrid1    &  Hybrid2 & Hybrid3           \\ \hline
\multirow{6}{*}{MNIST}         & SSLGuard        & 79.65\%          & 72.67\%          & 75.38\%          & 71.24\%   & 70.38\%   & 70.99\%             \\
                               & SSLGuard + BESA     & \textbf{97.24\%} & \textbf{96.23\%} & \textbf{97.09\%} & \textbf{95.87\%} & 
                               \textbf{94.66\%} &
                               \textbf{95.03\%} 
                                \\
                               & StolenEncoder   & 87.39\%          & 85.46\%          & 83.59\%          & 81.33\%          & 
                               81.32\%          &
                               82.01\%          
                                      \\
                               & StolenEncoder + BESA & \textbf{98.24\%} & \textbf{97.68\%} & \textbf{97.17\%} & \textbf{93.99\%} & 
                               \textbf{93.56\%} &
                               \textbf{92.29\%} 
                                \\
                               & Cont-Steal      & 90.71\%          & 88.44\%          & 91.25\%          & 87.54\%          & 
                               88.11\%          &
                               88.26\%          
                                    \\
                               & Cont-Steal + BESA    & \textbf{99.11\%} & \textbf{98.54\%} & \textbf{99.43\%} & \textbf{97.51\%} & 
                               \textbf{97.15\%} &
                               \textbf{96.13\%} 
                                \\ \hline
\multirow{6}{*}{Fashion-MNIST} & SSLGuard        & 77.21\%          & 73.24\%          & 72.54\%          & 68.45\% &  67.66\% & 68.88\%               \\
                               & SSLGuard + BESA      & \textbf{96.43\%} & \textbf{94.58\%} & \textbf{93.44\%} & \textbf{89.22\%} & 
                               \textbf{90.00\%} &
                               \textbf{89.68\%} 
                                \\
                               & StolenEncoder   & 85.49\%          & 84.25\%          & 83.98\%          & 77.45\%          & 
                               77.50\%          &
                               77.63\%          
                                       \\
                               & StolenEncoder + BESA & \textbf{97.55\%} & \textbf{94.18\%} & \textbf{95.46\%} & \textbf{91.32\%} & 
                               \textbf{91.89\%} &
                               \textbf{91.05\%} 
                                \\
                               & Cont-Steal      & 89.25\%          & 90.16\%          & 88.29\%          & 83.23\%          & 
                               84.33\%          &
                               85.01\%          
                                     \\
                               & Cont-Steal + BESA     & \textbf{97.58\%} & \textbf{96.84\%} & \textbf{93.43\%} & \textbf{93.98\%} & 
                               \textbf{93.27\%} &
                               \textbf{93.51\%} 
                                \\ \hline
\multirow{6}{*}{CIFAR-10}      & SSLGuard        & 62.98\%          & 62.49\%          & 63.45\%          & 61.11\% & 60.91\% & 61.51\%               \\
                               & SSLGuard + BESA      & \textbf{74.53\%} & \textbf{71.34\%} & \textbf{72.88\%} & \textbf{70.10\%} & 
                               \textbf{71.33\%} &
                               \textbf{69.35\%} 
                             \\
                               & StolenEncoder   & 63.55\%          & 64.57\%          & 62.33\%          & 61.91\%          & 
                               61.99\%          &
                               61.86\%          
                                     \\
                               & StolenEncoder + BESA & \textbf{76.12\%} & \textbf{77.23\%} & \textbf{78.59\%} & \textbf{77.05\%} & 
                               \textbf{76.84\%} &
                               \textbf{77.05\%} 
                               \\
                               & Cont-Steal      & 64.54\%          & 66.15\%          & 65.94\%          & 63.41\%          & 
                               62.87\%          &
                               63.24\%          
                                       \\
                               & Cont-Steal + BESA     & \textbf{78.92\%} & \textbf{76.33\%} & \textbf{75.20\%} & \textbf{73.22\%} & 
                               \textbf{73.59\%} &
                               \textbf{73.26\%} 
                               \\ \hline
\multirow{6}{*}{SVHN}          & SSLGuard        & 60.35\%          & 60.21\%          & 59.34\%          & 58.22\%   & 59.36\% & 59.01\%               \\
                               & SSLGuard + BESA      & \textbf{68.24\%} & \textbf{66.98\%} & \textbf{65.54\%} & \textbf{66.29\%} & 
                               \textbf{67.26\%} &
                               \textbf{66.81\%} 
                                \\
                               & StolenEncoder   & 61.24\%          & 60.87\%          & 62.45\%          & 60.98\%          & 
                               61.54\%          &
                               61.37\%          
                                      \\
                               & StolenEncoder + BESA & \textbf{69.25\%} & \textbf{70.11\%} & \textbf{68.82\%} & \textbf{66.45\%} & 
                               \textbf{67.35\%} &
                               \textbf{67.19\%} 
                                \\
                               & Cont-Steal      & 62.47\%          & 63.11\%          & 62.73\%          & 61.04\%          & 
                               62.05\%          &
                               61.99\%          
                                     \\
                               & Cont-Steal + BESA     & \textbf{70.47\%} & \textbf{70.98\%} & \textbf{71.47\%} & \textbf{72.08\%} & 
                               \textbf{73.18\%} &
                               \textbf{72.98\%} 
                                \\ \hline
\multirow{6}{*}{ImageNette}          & SSLGuard        & 70.68\%          & 68.32\%          & 69.34\%          & 55.20\% & 55.96\% & 56.32\%              \\
                               & SSLGuard + BESA      & \textbf{78.41\%} & \textbf{72.82\%} & \textbf{76.53\%} & \textbf{60.23\%} & 
                               \textbf{61.93\%} &
                               \textbf{62.07\%} 
                               \\
                               & StolenEncoder   & 69.42\%          & 67.21\%          & 68.11\%          & 63.01\%          & 
                               63.68\%          &
                               64.08\%          
                                     \\
                               & StolenEncoder + BESA & \textbf{74.25\%} & \textbf{72.13\%} & \textbf{75.66\%} & \textbf{70.15\%} & 
                               \textbf{71.16\%} &
                               \textbf{71.82\%} 
                               \\
                               & Cont-Steal      & 76.84\%          & 73.12\%          & 68.35\%          & 70.01\%          & 
                               71.13\%          &
                               71.96\%          
                                    \\
                               & Cont-Steal + BESA     & \textbf{80.72\%} & \textbf{78.67\%} & \textbf{72.67\%} & \textbf{76.77\%} & 
                               \textbf{77.25\%} &
                               \textbf{78.33\%} 
                               \\ \hline
                            
\end{tabular}%
}
\end{table*}

\section{Experiments}
\label{sec:expr}

\subsection{Experimental Settings}

\subsubsection{Datasets and Target Encoders}

We conducted experiments on four commonly used datasets: MNIST, Fashion-MNIST, CIFAR-10, SVHN, and ImageNette. We trained target encoders on these datasets using VGG16 and ResNet-34 respectively. For the contrastive learning algorithm, we employed three widely adopted methods: SimCLR, MoCo, and BYOL.

\subsubsection{State-of-the-Art (SOTA) Attacks}

We list three SOTA encoder stealing attacks improved by the use of BESA.

\textbf{SSLGuard} \cite{cong2022sslguard}. The encoder stealing approach in \textit{SSLGuard} leverages the samples in the shadow dataset to simply query the target encoder for optimizing the surrogate encoder.

\textbf{StolenEncoder} \cite{liu2022stolenencoder}. \textit{StolenEncoder} employs data augmentation to enhance the loss function for optimizing the surrogate encoder. Moreover, they leverage the inner characteristics of the pre-trained encoder to decrease the query budget.

\textbf{Cont-Steal} \cite{sha2023can}. Inspired by the concept of contrastive learning, \textit{Cont-Steal} ensures that the surrogate feature vector of an image is closely aligned with its target feature vector, while also creating a distinction between feature vectors of different images.

\subsubsection{Settings of BESA}

In this study, we implement BESA within three established encoder stealing attack frameworks, resulting in their enhanced variations. The experimental settings for existing attacks are selected based on the ones in their original paper for a fairer comparison, and the results are the average based on ten repeated experiments to avoid the occasional situation.

In order to ensure that the meta-classifiers can generalize effectively in accurately detecting the specific defense method used by the service provider, a substantial number of shadow encoders with various architectures (e.g., ResNet, VGG, etc.) are trained for the surrogate encoder. In our experiments, the default number of shadow encoders is set to 128.
All encoders undergo training utilizing contrastive algorithms that can be found on GitHub. For training the surrogate encoders, we use the following default hyperparameters across all experiments unless otherwise specified: a learning rate of 0.1, a batch size of 128, and the SGD optimizer with momentum 0.9. Each training run lasted for 100 epochs, and learning rate decay was applied with a factor of 0.1 every 60 epochs. Finally, we initialized the surrogate encoder with ResNet-50 as the architecture and used SimCLR for pre-training. For the experimental hardware platform, we utilized two NVIDIA 4090 GPUs, each equipped with 24 GB of memory.

\subsubsection{State-of-the-Art Defenses}

Based on the discussion in state-of-the-art works \cite{liu2022stolenencoder,sha2023can}, we here mainly concentrate on three commonly adopted perturbation-based defense methods: Top-K, rounding, and noise poisoning.

\begin{itemize}
    \item \textbf{Top-K.} The top-K algorithm selects and preserves the $K$ largest elements in the feature vector while setting all other elements to zero.
    \item \textbf{Rounding (RD).} Rounding is performed on the feature vector, preserving a specific number of digits after the decimal point for each element.
    \item \textbf{Noise Poisoning (NP).} NP adding Gaussian noise to the feature vector. The default setting for $z \in \mathcal{N}(0, \sigma^{2})$ is $\sigma^{2} = 0.2$.
\end{itemize}


\subsection{Experimental Results}

\subsubsection{Performance of BESA}
\label{sec:expr-impr}

Table \ref{tab:enhance} illustrates the experimental results of BESA for state-of-the-art encoder stealing attacks under three perturbation-based defenses: Top-k, rounding, and noise poisoning. The analysis of experimental results is explained in detail below.

\textbf{Single Defense:} Our proposed BESA effectively enhances existing canonical encoder stealing attacks for all three cases of the single defense listed. We have observed that the defenses significantly reduce the accuracy of the surrogate encoder, particularly in the earlier attack in \textit{SSLGuard}. When employing perturbation-based defense methods, BESA improves the performance by up to 24.63\% compared to state-of-the-art encoder stealing attacks. While the improvement on existing attacks may be decreased for more complex datasets (e.g., ImageNette, SVHN), the surrogate encoder is still able to enhance their accuracy by up to 16.26\%.

When employing hybrid defenses, such as the combination of random dropout (RD), noise perturbation (NP), and top-k suppression, service providers typically achieve stronger protection than using a single defense alone. To reflect realistic scenarios where the defense strategy may change dynamically across queries, we adopt a per-query perturbation detection approach in BESA. We evaluate BESA under three hybrid defense settings—RD+NP (Hybrid1), RD+Top-K (Hybrid2), and NP+Top-K (Hybrid3)—as summarized in Table I. Although these hybrid strategies significantly increase the difficulty of encoder stealing, BESA consistently improves the surrogate encoder performance across all datasets and attack baselines. For instance, under Hybrid3 on MNIST, BESA enhances the surrogate accuracy of \textit{SSLGuard} from 70.99\% to 92.29\%. These results demonstrate that the perturbation detection module in BESA can reliably identify and handle complex, mixed defenses with high accuracy, maintaining its effectiveness even under more aggressive protection schemes.

\begin{figure*}[tb]
  \subfloat[MNIST]{\includegraphics[width=0.25\textwidth]{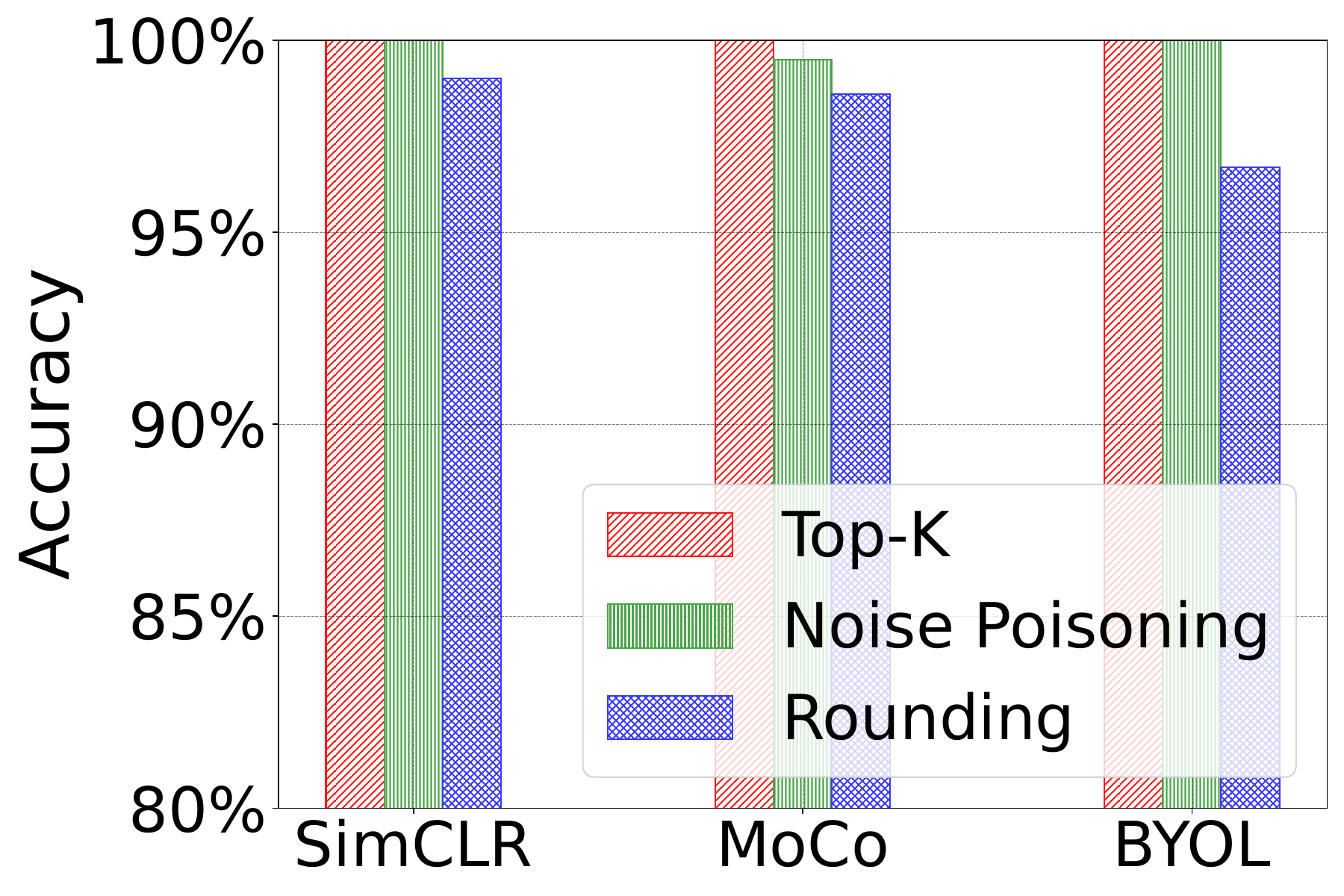}\label{fig:mnist-d}}
 \hfill 	
  \subfloat[Fashion-MNIST]{\includegraphics[width=0.25\textwidth]{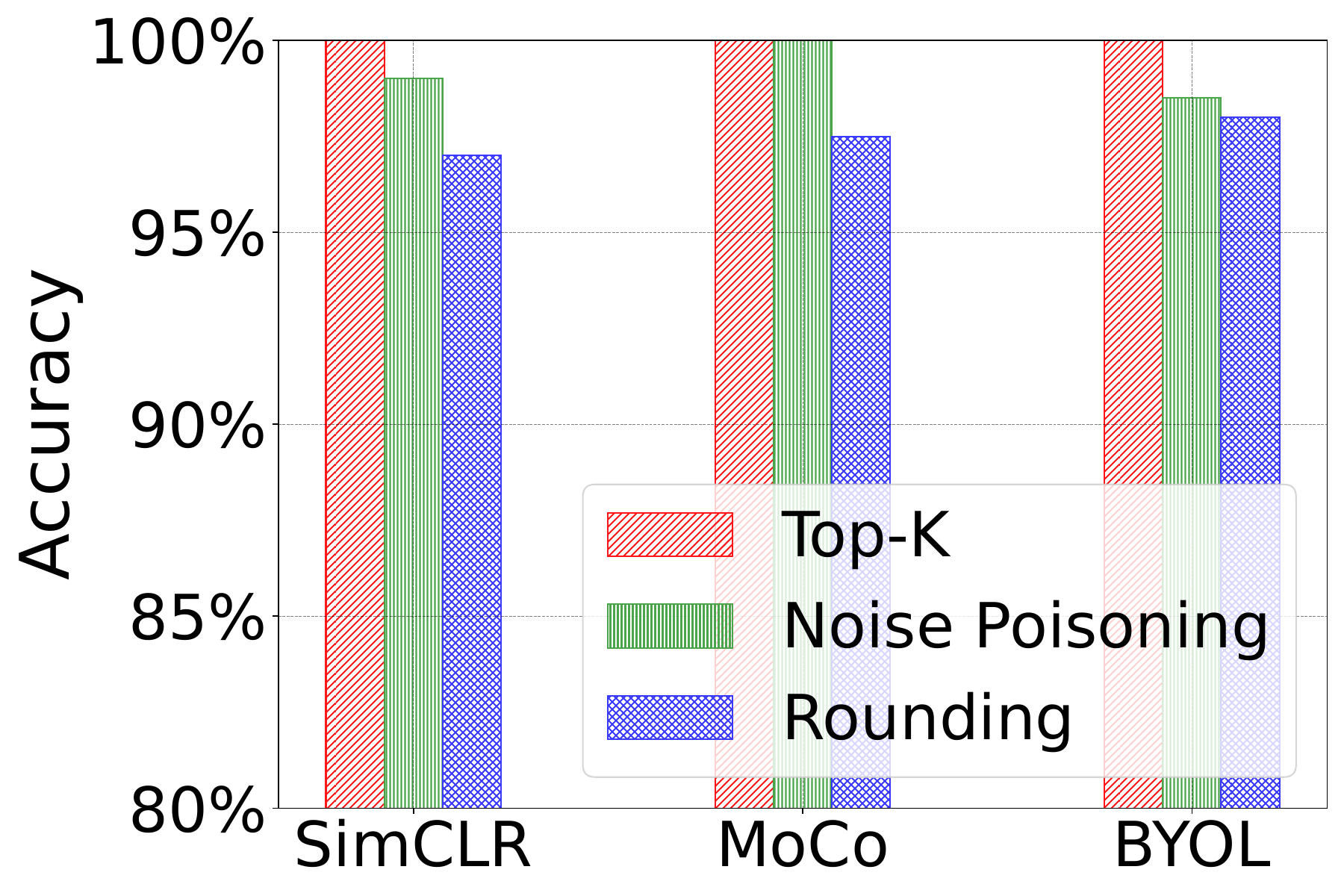}\label{fig:fmnist-d}}
 \hfill	
  \subfloat[CIFAR-10]{\includegraphics[width=0.25\textwidth]{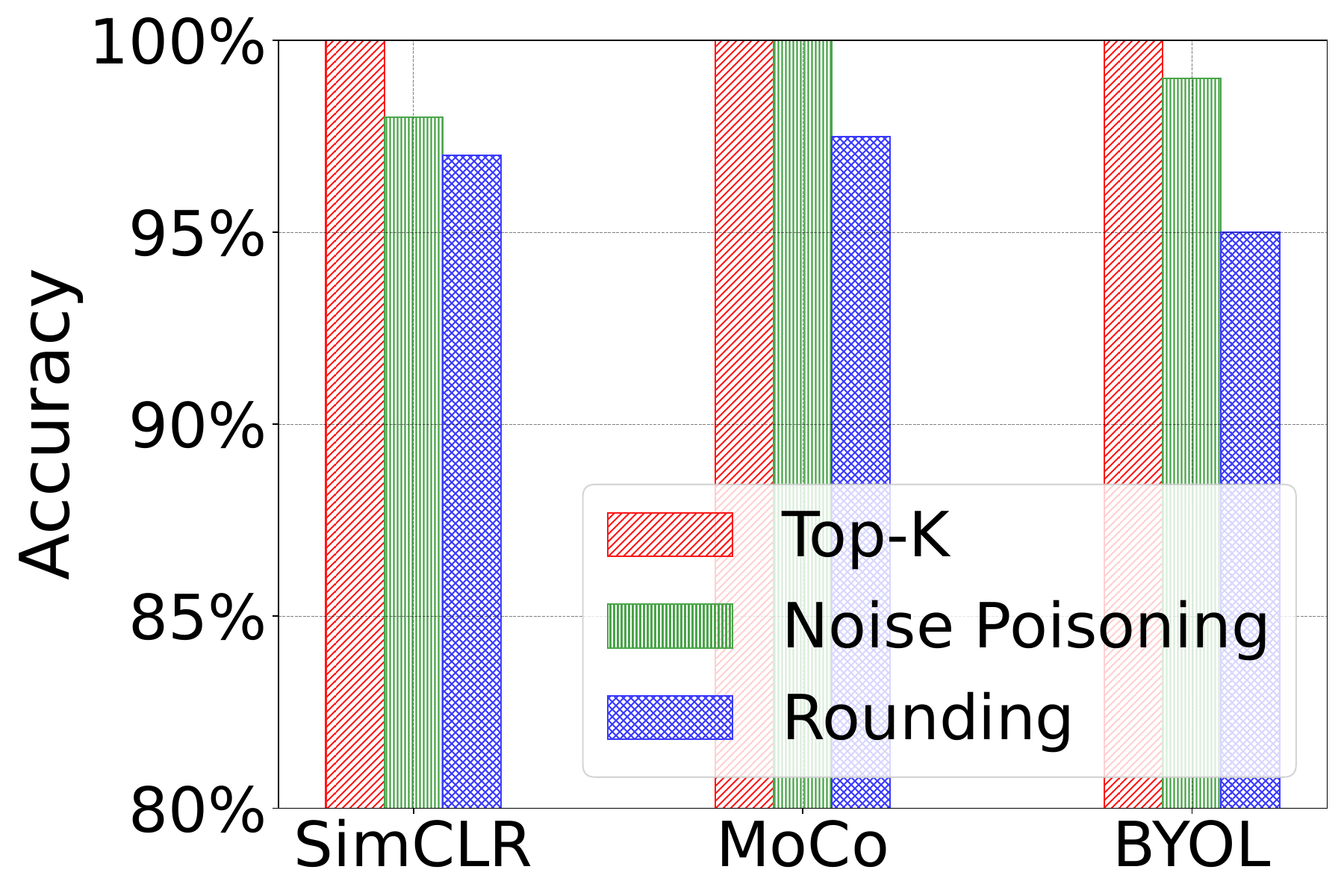}\label{fig:cifar-d}}
 \hfill	
  \subfloat[SVHN]{\includegraphics[width=0.25\textwidth]{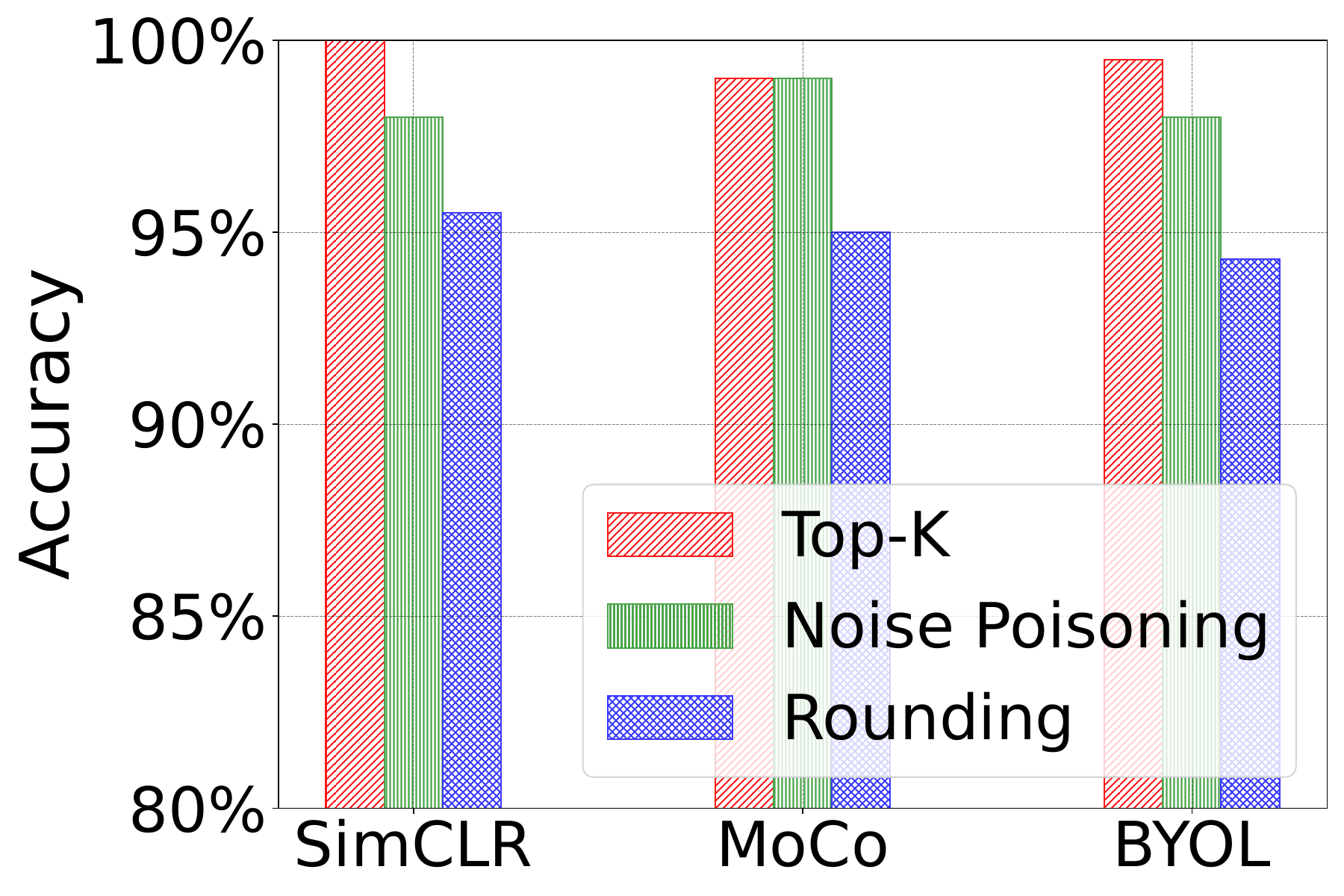}\label{fig:svhn-d}}
\caption{Detection accuracy of meta-classifiers in BESA.}
\label{fig:detect}
\end{figure*}

\begin{table*}[]
\centering
\caption{ACCURACY OF THE SURROGATE ENCODER FOR DIFFERENT DATASETS on different Unknown Defenses.}
\label{tab:unknown}
\resizebox{10cm}{!}{
\begin{tabular}{cc|ccc}
\hline
Dataset                        & Method               & Un-Top-k & Un-RD   & Un-NP   \\ \hline
\multirow{6}{*}{MINST}         & SSLGuard             & 57.43\%  & 64.94\% & 69.44\% \\
                               & SSLGuard + BESA      & \textbf{64.47\%}  & \textbf{71.32\%} & \textbf{86.98\%} \\
                               & StolenEncoder        & 62.35\%  & 68.14\% & 78.28\% \\
                               & StolenEncoder + BESA & \textbf{70.88\%}  & \textbf{75.72\%} & \textbf{85.45\%} \\
                               & Cont-Steal           & 65.71\%  & 75.10\% & 81.23\% \\
                               & Cont-Steal + BESA    & \textbf{72.36\%}  & \textbf{79.50\%} & \textbf{88.60\%} \\
                               \hline
\multirow{6}{*}{Fashion-MNIST} & SSLGuard             & 62.22\%  & 61.86\% & 67.34\% \\
                               & SSLGuard + BESA      & \textbf{68.57\%}  & \textbf{73.37\%} & \textbf{84.91\%}  \\
                               & StolenEncoder        & 59.96\%  & 66.52\% & 74.83\% \\
                               & StolenEncoder + BESA & \textbf{70.39\%}  & \textbf{71.24\%} & \textbf{82.11\%} \\
                               & Cont-Steal           & 68.04\%  & 73.65\% & 80.65\% \\
                               & Cont-Steal + BESA    & \textbf{75.00\%}  & \textbf{81.41\%} & \textbf{87.33\%} \\
                               \hline
\multirow{6}{*}{CIFAR-10}      & SSLGuard             & 63.25\%  & 68.29\% & 60.34\% \\
                               & SSLGuard + BESA      & \textbf{72.96\%}  & \textbf{77.34\%} & \textbf{65.17\%} \\
                               & StolenEncoder        & 61.82\%  & 59.33\% & 60.66\% \\
                               & StolenEncoder + BESA & \textbf{69.13\%}  & \textbf{71.62\%} & \textbf{71.24\%} \\
                               & Cont-Steal           & 58.37\%  & 65.29\% & 62.93\% \\
                               & Cont-Steal + BESA    & \textbf{68.38\%}  & \textbf{73.06\%} & \textbf{70.45\%} \\
                               \hline
\multirow{6}{*}{SVHN}          & SSLGuard             & 68.25\%  & 57.20\% & 57.36\% \\
                               & SSLGuard + BESA      & \textbf{76.59\%}  & \textbf{63.01\%} & \textbf{65.77\%} \\
                               & StolenEncoder        & 59.63\%  & 58.11\% & 59.87\% \\
                               & StolenEncoder + BESA & \textbf{65.00\%}  & \textbf{67.26\%} & \textbf{66.17\%} \\
                               & Cont-Steal           & 63.03\%  & 65.73\% & 60.99\% \\
                               & Cont-Steal + BESA    & \textbf{72.90\%}  & \textbf{71.25\%} & \textbf{68.33\%} \\
                               \hline
\multirow{6}{*}{ImageNette}    & SSLGuard             & 55.21\%  & 54.99\% & 52.63\% \\
                               & SSLGuard + BESA      & \textbf{65.37\%}  & \textbf{67.82\%} & \textbf{55.78\%} \\
                               & StolenEncoder        & 56.61\%  & 52.34\% & 59.87\% \\
                               & StolenEncoder + BESA & \textbf{62.38\%}  & \textbf{60.84\%} & \textbf{66.17\%} \\
                               & Cont-Steal &
                               58.35\% & 55.60\% & 60.83\% \\
                               & Cont-Steal + BESA    & \textbf{65.39\%}  & \textbf{72.09\%} & \textbf{68.33\%} \\ \hline
\end{tabular}}
\end{table*}

\textbf{Unknown Defense:} To evaluate the robustness of BESA under unknown defense strategies, we simulate scenarios where the attacker trains BESA using a subset of known defenses, then applies it to encoders protected by unseen defenses. Previously, we considered a setting where BESA is trained on Top-K and rounding, and tested on NP (denoted as Un-NP). We now extend this evaluation to two additional settings: (1) training on NP and RD but testing on Top-K (Un-Top-k), and (2) training on Top-K and NP but testing on RD (Un-RD). As shown in Table~\ref{tab:unknown}, BESA consistently improves the accuracy of surrogate encoders across all datasets, even under these unseen defenses. For example, under Un-Top-k on MNIST, the surrogate encoder from \textit{SSLGuard + BESA} achieves 64.47\%, significantly higher than the baseline SSLGuard (57.43\%). These results demonstrate that BESA retains a strong degree of generalization and transferability, making it effective even in the presence of novel or unseen perturbation-based defenses.

\subsubsection{Performance on Perturbation Detection}
\label{sec:expr-detection}

In this subsection, we assess how well the meta-classifier can detect the defensive techniques used by the service provider. Following the training of the meta-classifiers, we assess their prediction accuracy by testing on an additional 128 shadow encoders, each protected by distinct defense strategies.

From Fig. \ref{fig:detect}, the results indicate that the meta-classifiers achieve an accuracy of over 98\% for all three contrastive algorithms utilized in pre-training the target encoder, particularly for typical datasets like MNIST and FashionMNIST. However, the performance of the meta-classifiers exhibits a slight decline when dealing with complex datasets, such as CIFAR-10 and SVHN. Notably, the presence of the \textit{Rounding} defense can be identified by examining the number of digits in the feature vectors. Interestingly, both \textit{Top-K} and \textit{Noise Poisoning} defense approaches can be accurately detected with almost perfect accuracy of nearly 100\%, possibly due to distinguishing characteristics present in the feature vectors, such as zero values or added noise.

\subsection{Impact Factors}
\label{sec:ablation}

\subsubsection{Architecture of the Surrogate Encoder}

In our previous experiments, we trained the surrogate encoder using the complex architecture of ResNet-50. 
This choice is motivated by the belief that a more capable architecture will better emulate the functionality of the target encoder.
In this part, we aim to investigate the impact of different architectural choices for surrogate encoders on attack performance. Specifically, when using VGG16 as the target encoder for CIFAR-10 and ResNet-34 for SVH, we select various architectures for the surrogate encoder. In the case of CIFAR-10, the surrogate encoder options consisted of AlexNet, ResNet-18, ResNet-34, and VGG16, while for SVHN, the options included AlexNet, ResNet-18, ResNet-34, and VGG16. 

\begin{figure}[htbp]
  \subfloat[CIFAR-10]{\includegraphics[width=0.5\columnwidth]{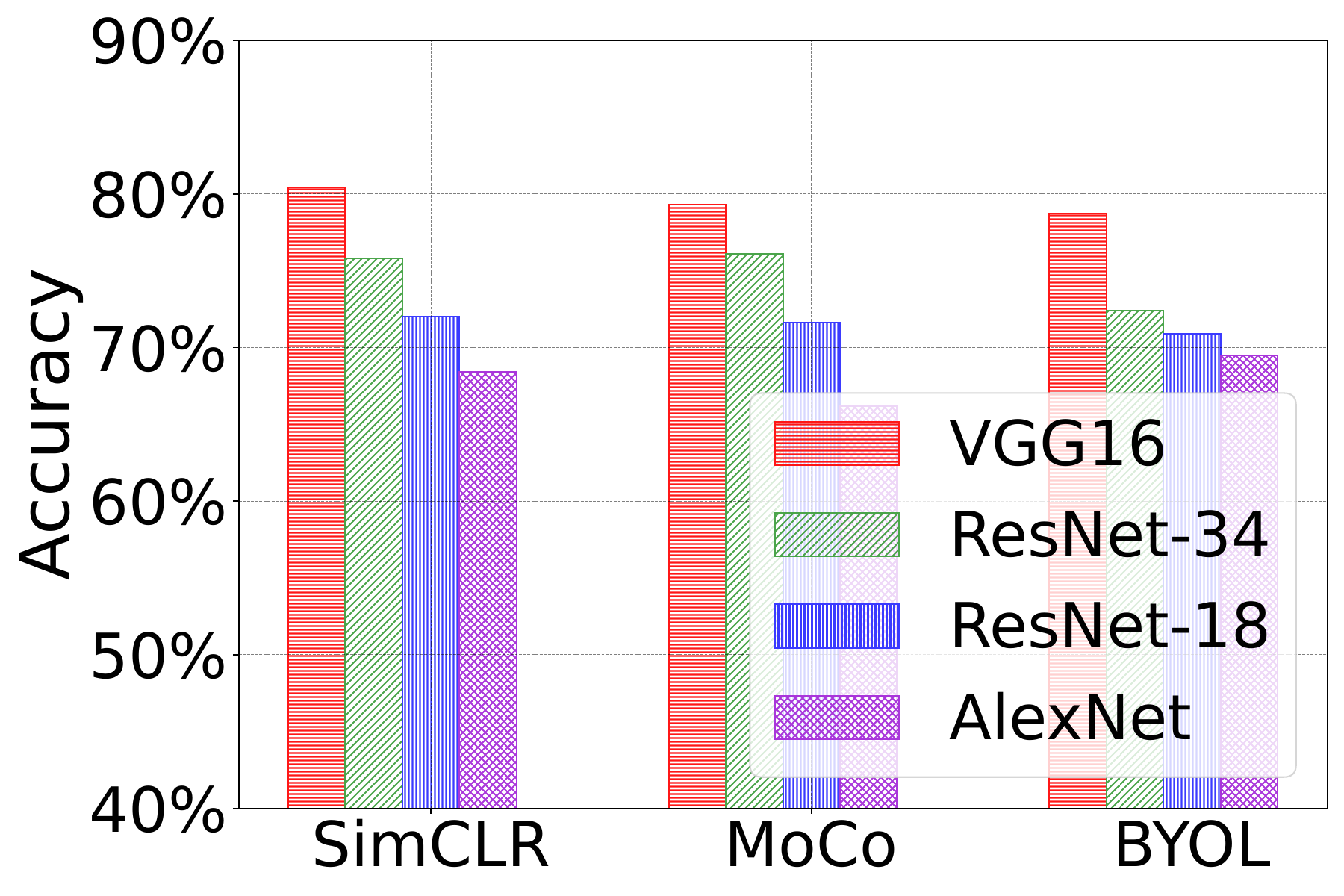}\label{fig:cifar-a}}
 \hfill	
  \subfloat[SVHN]{\includegraphics[width=0.5\columnwidth]{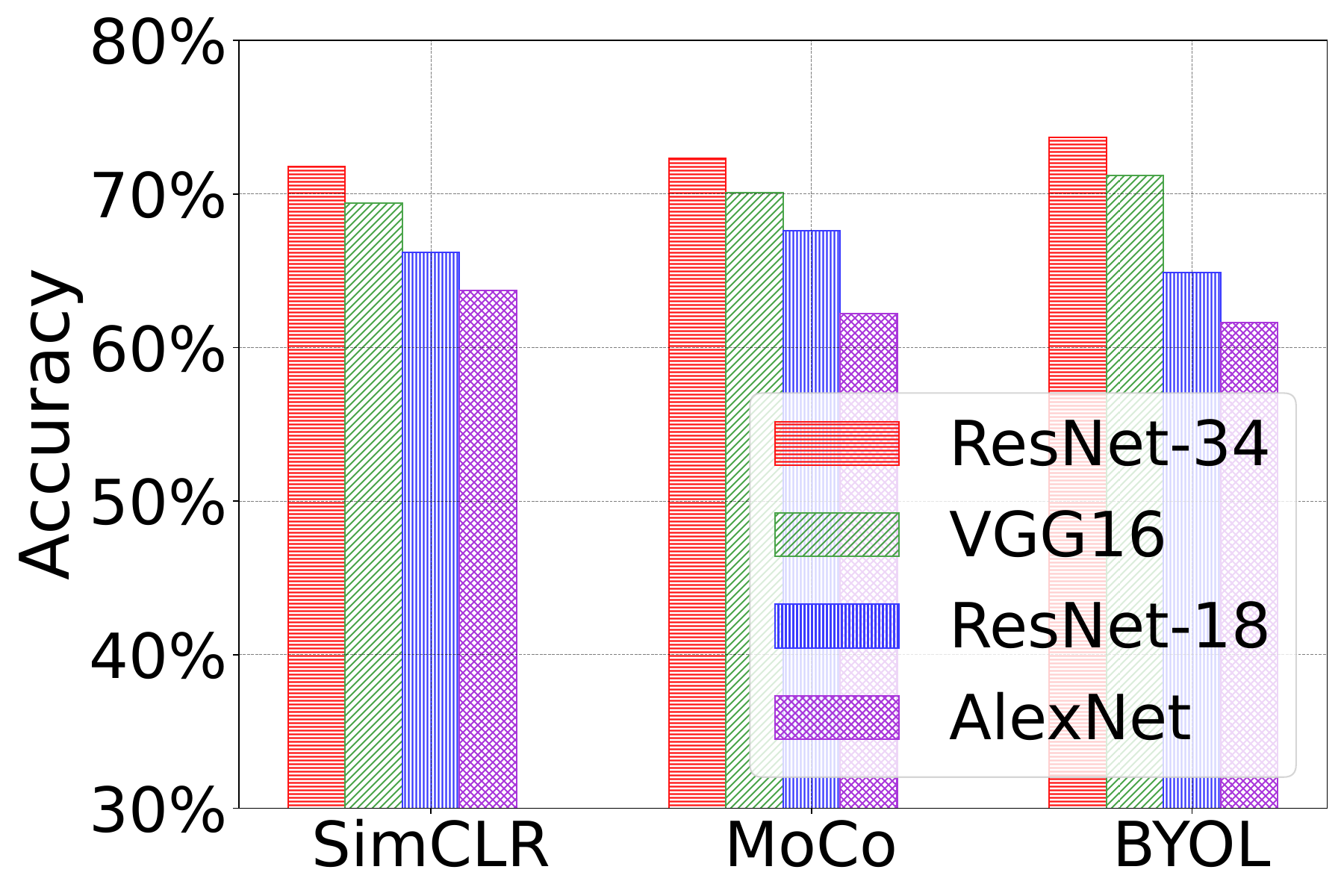}\label{fig:svhn-a}}
\caption{Impact of architecture choice on BESA accuracy.}
\label{fig:arch}
\end{figure}

As shown in Fig. \ref{fig:arch}, our results indicate that it is possible to successfully steal the target encoder even when there is a mismatch between its architecture and that of the surrogate encoder, provided that the surrogate encoder's architecture is complex enough. Moreover, our findings suggest that a more intricate architecture for the surrogate encoder enhances its performance by enabling it to effectively mimic the target encoder through utilizing a greater amount of information.

\subsubsection{Loss functions}
\label{sec:ablation-loss}

The core element of the perturbation recovery in BESA involves minimizing the distance between the perturbed feature vectors and the unperturbed ones. Notably, the main objective of BESA is to bypass the perturbation-based defense mechanisms by providing recovered clean feature vectors. Therefore, selecting a more suitable loss function can significantly enhance BESA performance. In this study, we evaluate the effects of three distinct loss functions: cosine similarity, $\mathcal{L}_{2}$ norm, and $\mathcal{L}_{1}$ norm. 

\begin{figure}[htbp]
  \subfloat[MNIST]{\includegraphics[width=0.5\columnwidth]{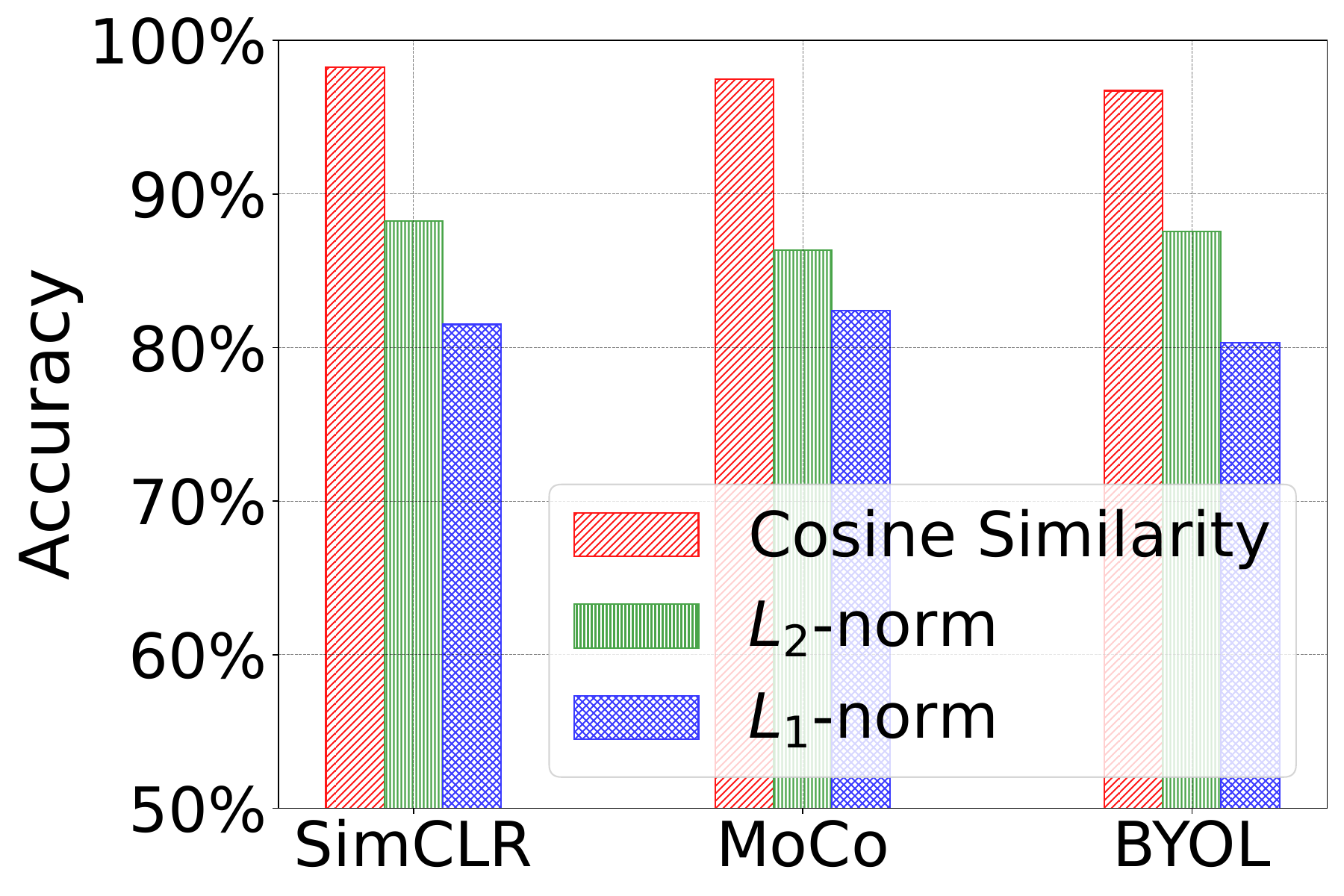}\label{fig:mnist-di}}
 \hfill	
  \subfloat[CIFAR-10]{\includegraphics[width=0.5\columnwidth]{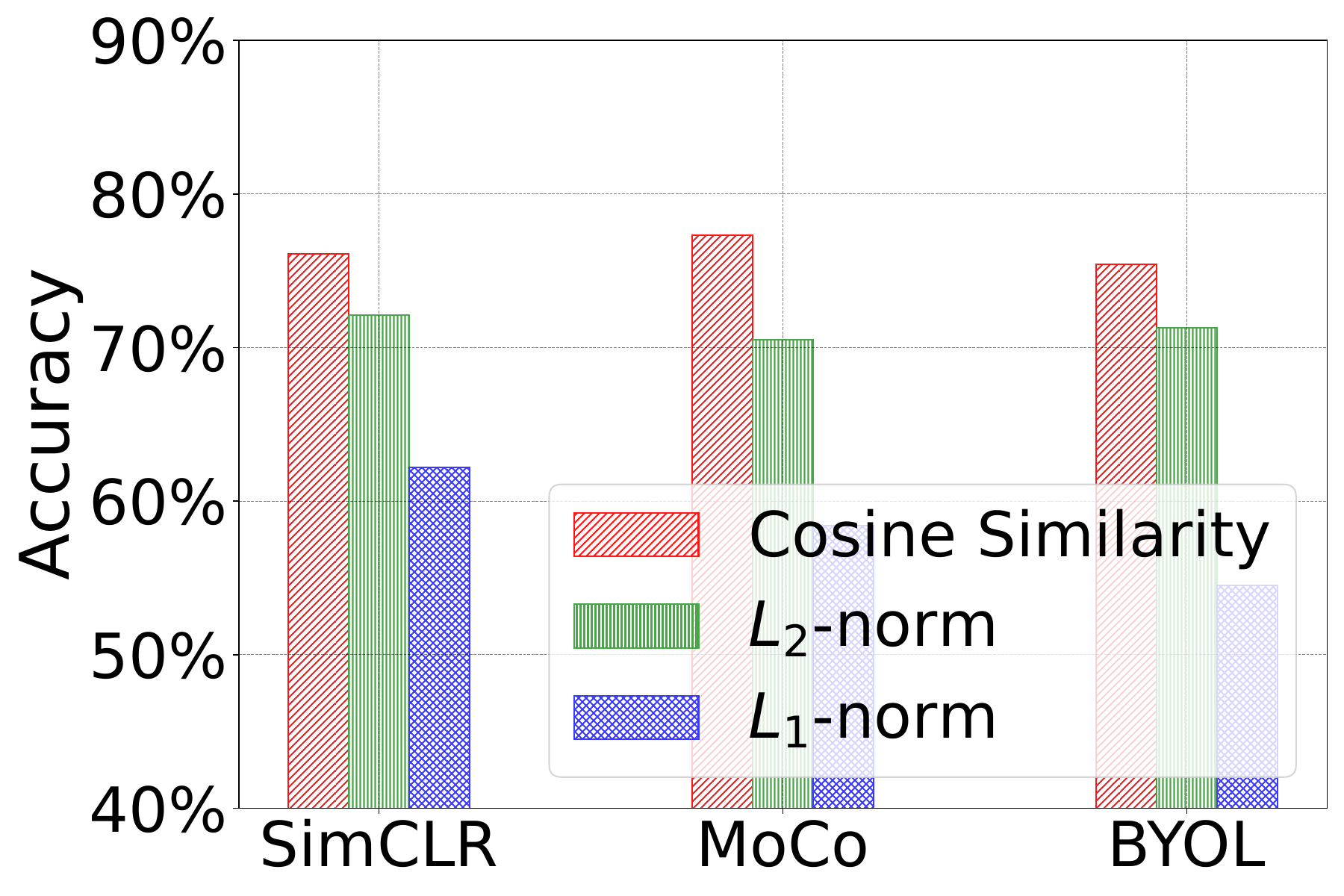}\label{fig:cifar-di}}
\caption{Impact of loss functions on BESA accuracy.}
\label{fig:dist}
\end{figure}

Fig. \ref{fig:dist} illustrates the outcomes associated with each loss function. In both the MNIST and CIFAR-10 datasets, cosine similarity demonstrates superior performance compared to the other two loss functions. One potential rationale behind this is that cosine similarity more effectively captures the resemblance between distinct feature vectors in contrast to the other functions \cite{nguyen2022survey}. In comparing the efficacy of $\mathcal{L}_2$ distance with $\mathcal{L}_1$ distance, it was observed that $\mathcal{L}_2$ distance delivers enhanced accuracy. This is likely due to the fact that in $\mathcal{L}_2$ distance, the differences between the feature vectors generated by the target encoder and the surrogate encoder are minimized across all dimensions.

\subsection{Time Costs}

\begin{table}[]
\caption{Training time OF BESA (SECONDS)}
\label{tabII}
\begin{tabular}{c|ccc}
\hline
Dataset       & Shadow Model & Meta- Classifier & Generator \\ \hline
MINST         & 13           & 210              & 15326     \\
Fashion-MNIST & 15           & 305              & 17025     \\
CIFAR-10      & 35           & 272              & 20985     \\
SVHN          & 62           & 291              & 31025     \\
ImageNette    & 52           & 1650             & 28650     \\ \hline
\end{tabular}
\end{table}

The computational cost of BESA can be separated into two stages: the offline training phase, where shadow models, meta-classifiers, and generators are built; and the online inference phase, where these trained components are used to detect and recover perturbed outputs during encoder stealing. The offline training introduces a one-time cost. As shown in Table~\ref{tabII}, we report the training time across four benchmark datasets. The time varies depending on the dataset complexity and the number of shadow models involved. Notably, although SVHN requires more shadow encoders, the overall training time remains manageable and is performed only once. In cases where a new perturbation defense is encountered and known to the attacker, the corresponding components can be trained offline and integrated into the existing framework with minimal disruption. Moreover, the trained meta-classifiers and generators are reusable across different target encoders, enabling multiple encoder stealing attempts without retraining.

For unknown or novel defenses not previously modeled, our results show that BESA exhibits a degree of robustness, likely due to shared patterns between new and existing defense strategies. This allows for partial generalization even without retraining. During the online attack phase, BESA only performs lightweight inference using the pre-trained meta-classifier and generator, resulting in negligible overhead when integrated into standard encoder stealing pipelines.

\section{Discussion}
In this section, we discuss the limitations and ethical problems of BESA.
\subsection{Ethical Problem}
In this paper, we propose the BESA to enhance existing encoder stealing attacks and show the inadequacy of existing perturbation defenses in the face of such attacks. While our work reveals the limitations of current defenses, we emphasize that the BESA is only a further exploration of existing attack methods, rather than advocating malicious use. We call on academia and industry to propose more effective defense strategies to protect pre-trained encoders from such attacks.

One potential mitigation is to limit the effectiveness of such attacks through request detection mechanisms. Since BESA attacks require a large number of query requests, the probability of successful attacks can be reduced by limiting the query frequency of each user and combining anomaly detection algorithms to identify and block malicious query behaviors. In addition, service providers can introduce stronger authentication mechanisms and behavioral analysis to strengthen defenses. Such measures can effectively prevent malicious use of the BESA and ensure normal use by legitimate users.

\subsection{Compatibility with Other Defense}
BESA is designed primarily to bypass perturbation-based defenses by recovering clean feature vectors from intentionally disrupted outputs. However, its design does not directly target detection-based or differential privacy (DP)-based defenses. Detection-based approaches typically analyze the distribution of query inputs to identify potential malicious behavior. Since BESA operates only on the returned feature vectors and does not control the query generation process, it may be affected by detection mechanisms depending on how queries are issued. That said, many detection-based systems return perturbed outputs rather than reject responses altogether, in which case BESA remains effective. To fully bypass such defenses, future extensions could incorporate more advanced query synthesis strategies that generate queries mimicking normal usage patterns. For DP-based defenses, their evaluation remains limited due to the lack of publicly available implementations. Once such mechanisms become accessible, we plan to extend our evaluation accordingly.

\subsection{Generalization}
The effectiveness of BESA's perturbation detection module relies on the meta-classifier’s ability to recognize defense-specific patterns in feature vectors. This requires training with outputs from shadow encoders equipped with various defenses. Because the attacker does not have access to the target encoder’s architecture, these shadow models must be diverse to ensure generalization. While this strategy works well in practice, it does come with additional training overhead. A potential future direction is to adopt meta-learning techniques that improve the adaptability of the meta-classifier using fewer shadow encoders. Additionally, as discussed in Section III, BESA operates under a black-box setting where the attacker has no access to model internals or original training data but can query the encoder and simulate common defenses. These assumptions reflect practical constraints in real-world Encoder-as-a-Service scenarios.

\section{Conclusion}
\label{sec:conclusion}

We propose a novel encoder stealing attack called BESA which allows the construction of a well-performing surrogate encoder, even in the presence of perturbation-based defenses protecting the target encoder. We identify the specific defense method utilized by the service provider using meta-classifiers and restore perturbed feature vectors using a generative model. Extensive experimental results demonstrate that BESA significantly enhances the accuracy of surrogate encoders when facing various defensive mechanisms.

\bibliography{ref.bib}
\bibliographystyle{IEEEtran}

\begin{IEEEbiography}
    [{\includegraphics[width=1in,height=1.25in, clip, keepaspectratio]{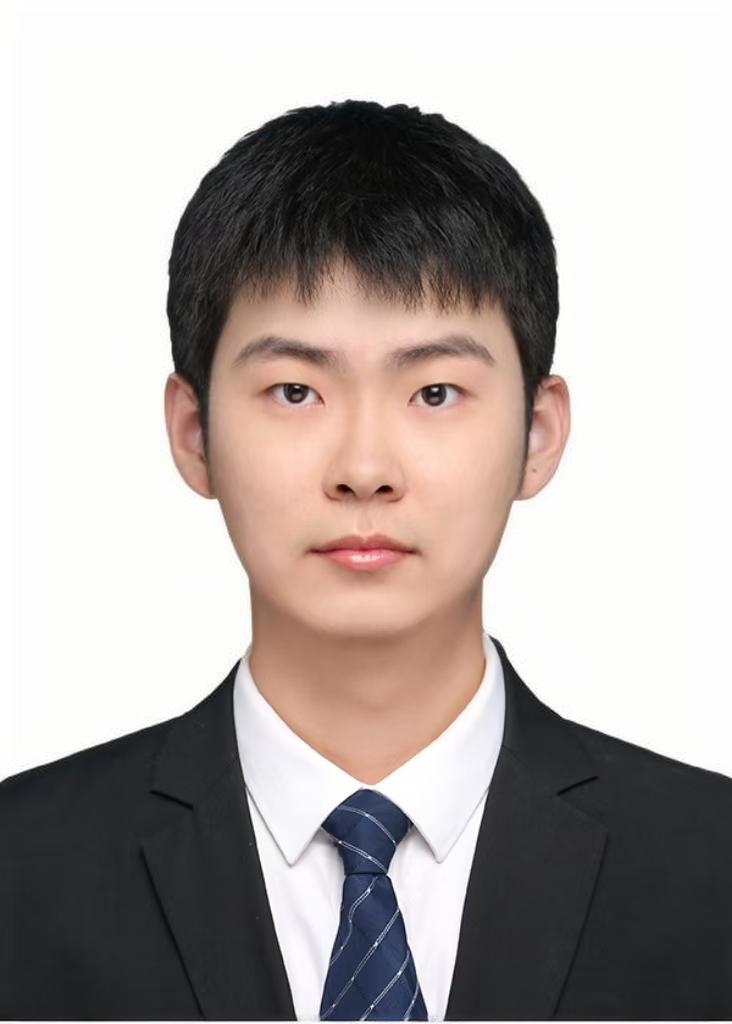}}]{Xuhao Ren}
    received his B.S. degree in the School of Information Engineering of Sichuan Agricultural University, Sichuan, China, in 2022. He is currently a master’s student in the School of Cyberspace Science and Technology, Beijing Institute of Technology. His research interests include applied cryptography.
\end{IEEEbiography}

\begin{IEEEbiography}
    [{\includegraphics[width=1in,height=1.25in, clip, keepaspectratio]{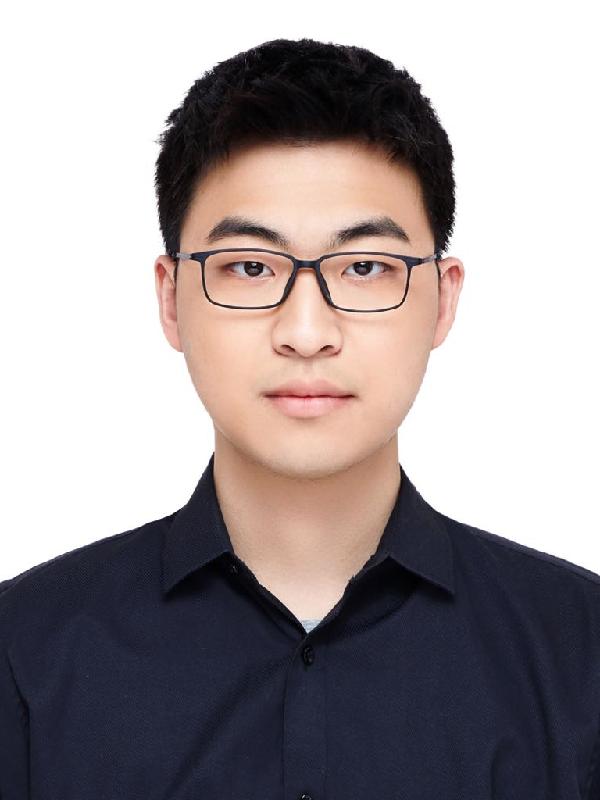}}]{Haotian Liang}
     received the B.S. degree from Lanzhou University in 2022. He is currently pursuing the master’s degree with the School of Cyberspace Science and Technology, Beijing Institute of Technology. His research interests include machine learning security, Internet of Things security, and cloud security.
\end{IEEEbiography}

\begin{IEEEbiography}
    [{\includegraphics[width=1in,height=1.25in, clip, keepaspectratio]{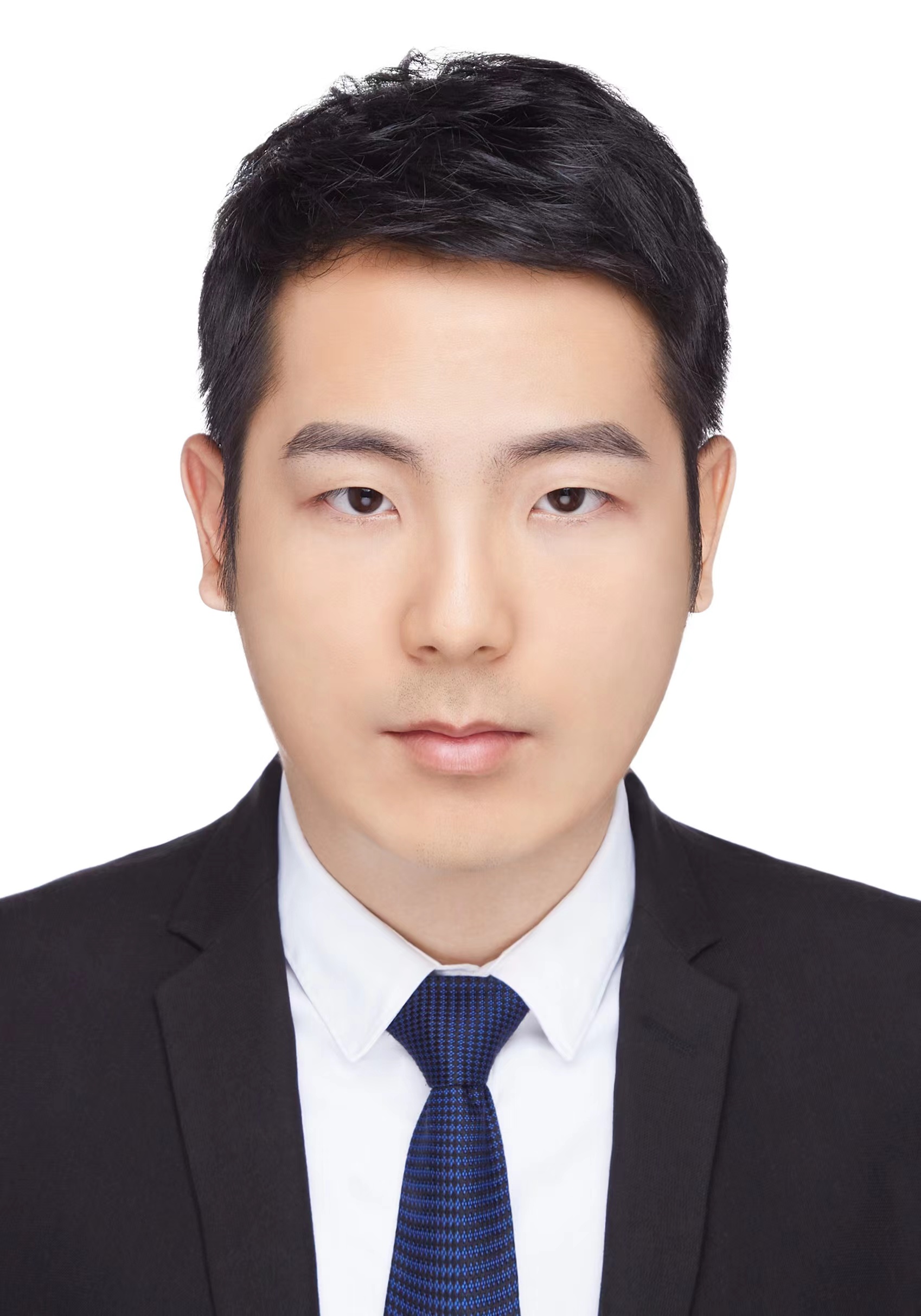}}]{Yajie Wang}
    received his Ph.D. degree in the School of Computer Science, Beijing Institute of Technology, Beijing, China. He is currently a postdoctoral at the School of Cyberspace Science and Technology, Beijing Institute of Technology. His main research interests include the robustness and vulnerability of artificial intelligence, cyberspace security, etc.
\end{IEEEbiography}

\begin{IEEEbiography}[{\includegraphics[width=1in,height=1.25in,clip,keepaspectratio]{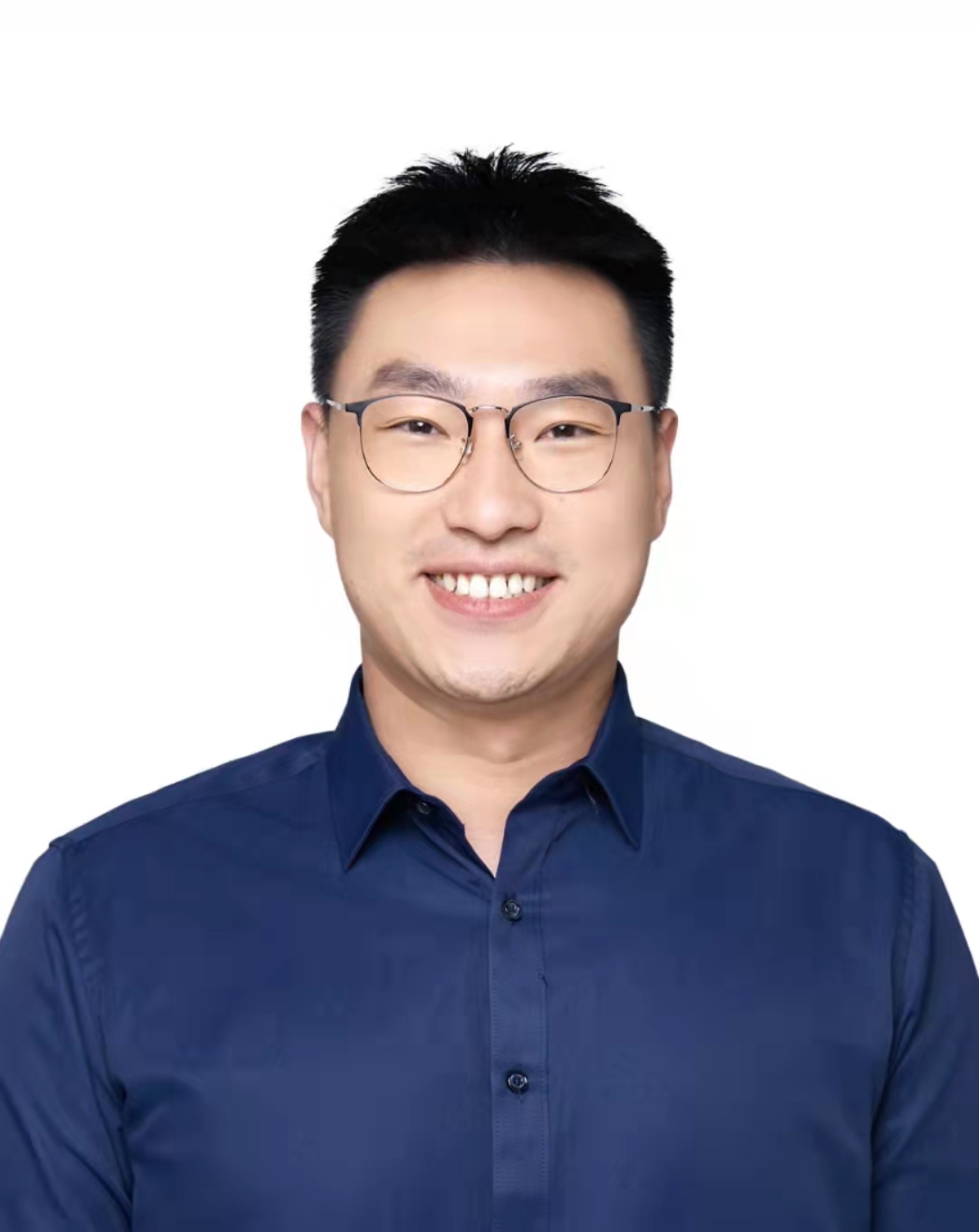}}]{Chuan Zhang} (Member, IEEE)
  received his Ph.D. degree in computer science from Beijing Institute of Technology, Beijing, China, in 2021. 
  From Sept. 2019 to Sept. 2020, he worked as a visiting Ph.D. student with the BBCR Group, Department of Electrical and Computer Engineering, University of Waterloo, Canada. He is currently an assistant professor at the School of Cyberspace Science and Technology, Beijing Institute of Technology. 
  His research interests include applied cryptography, machine learning, and blockchain.
\end{IEEEbiography}

\begin{IEEEbiography}
    [{\includegraphics[width=1in,height=1.25in, clip, keepaspectratio]{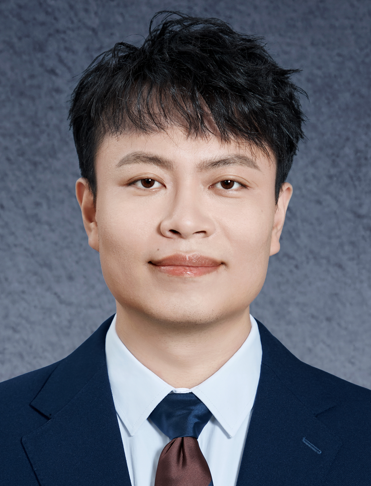}}]{Zehui Xiong} (Senior Member, IEEE) is currently an Assistant Professor at Singapore University of Technology and Design, and also an Honorary Adjunct Senior Research Scientist with Alibaba-NTU Singapore Joint Research Institute, Singapore. He received the PhD degree in Nanyang Technological University (NTU), Singapore. He was the visiting scholar at Princeton University and University of Waterloo. His research interests include wireless communications, Internet of Things, blockchain, edge intelligence, and Metaverse.
\end{IEEEbiography}

\begin{IEEEbiography}[{\includegraphics[width=1in,height=1.25in,clip,keepaspectratio]{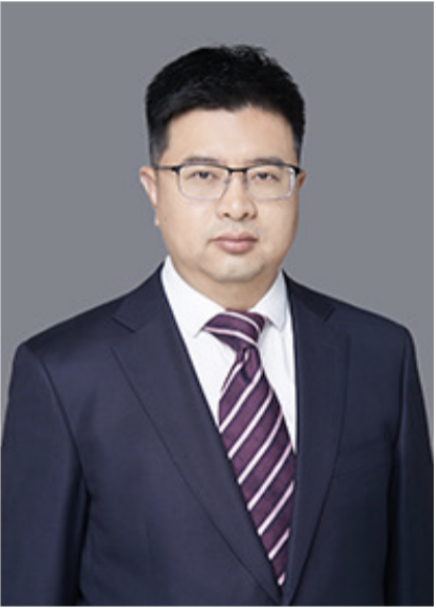}}]{Liehuang Zhu}
  (Senior Member, IEEE) received his Ph.D. degree in computer science from Beijing Institute of Technology, Beijing, China, in 2004. 
  He is currently a professor at the School of Cyberspace Science and Technology, Beijing Institute of Technology. 
  His research interests include security protocol analysis and design, group key exchange protocols, wireless sensor networks, and cloud computing.
\end{IEEEbiography}

\vfill

\end{document}